\documentclass[%
 reprint,
superscriptaddress,
 amsmath,amssymb,
 aps,
 prl,
floatfix,
]{revtex4-2}

\usepackage[mode=buildnew]{standalone}
\usepackage{tikz}
\usepackage{bm}
\usepackage{physics}
\usepackage{xcolor}
\usepackage{mathrsfs}

\usepackage{graphicx}
\usepackage{dcolumn}
\usepackage{bm}
\usepackage[normalem]{ulem}
\preprint{APS/123-QED}

\newcommand{\cbvn}{C$_\mathrm{B}$V$_\mathrm{N}$}
\newcommand{\ctcn}{C$_2$C$_\mathrm{N}$}

\begin{document}
\preprint{APS/123-QED}

\title{Signatures of Non-Markovianity in Cavity-QED with Color Centers in 2D Materials}
\author{Mark Kamper Svendsen}
\affiliation{CAMD, Department of Physics, Technical University of Denmark, 2800 Kgs. Lyngby,
Denmark}
\author{Sajid Ali}
\affiliation{CAMD, Department of Physics, Technical University of Denmark, 2800 Kgs. Lyngby,
Denmark}

\author{Nicolas Stenger}
\affiliation{Department of electrical and photonics engineering, Technical University of Denmark, 2800 Kgs. Lyngby, Denmark
}
\affiliation{
Center for Nanostructured Graphene, Technical University of Denmark, 2800 Kgs. Lyngby, Denmark
}
\affiliation{
NanoPhoton -- Center for Nanophotonics, Technical University of Denmark, 2800 Kgs. Lyngby, Denmark
}
\author{Kristian Sommer Thygesen}
\affiliation{CAMD, Department of Physics, Technical University of Denmark, 2800 Kgs. Lyngby,
Denmark}
\affiliation{
Center for Nanostructured Graphene, Technical University of Denmark, 2800 Kgs. Lyngby, Denmark
}
\author{Jake Iles-Smith}
  \email[Email: ]{jake.iles-smith@manchester.ac.uk}
 \affiliation{Department of Physics and Astronomy, The University of Manchester, Oxford Road, Manchester M13 9PL, United Kingdom}
  \affiliation{Department of Electrical and Electronic Engineering, The University of Manchester, Oxford Road, Manchester M13 9PL, United Kingdom}
\date{January 2022}

\begin{abstract}
Light-matter interactions of defects in two dimensional materials are expected to be profoundly impacted by strong coupling to phonons.
In this work, we combine \emph{ab initio} calculations of a defect in hBN, with a fully quantum mechanical and numerically exact description of a cavity-defect system to elucidate this impact.
We show that even at weak light-matter coupling, the dynamical evolution of the cavity-defect system has clear signatures of non-markovian phonon effects, and that linear absorption spectra show the emergence of hybridised light-matter-phonon states in regimes of strong light-matter coupling.
We emphasise that our methodology is general, and can be applied to a wide variety of material/defect systems.
\end{abstract}
\maketitle

Single-photon emission has been observed from a broad range of two dimensional materials (2DM)~\cite{Chakraborty2015wse2,He2015wse2,Koperski2015wse2,Srivastava2015wse2,klein2019site, michaelis2022single,kianinia2022quantum}.
Of particular interest to quantum technologies is the emission from defect complexes in hexagonal Boron Nitride (hBN)~\cite{Tran2016hbn,aharonovich2017quantum, fischer2021controlled,mendelson2021identifying,gottscholl2020initialization,hoese2020mechanical}, owing to the materials wide band-gap~\cite{cassabois2016hexagonal} and stability of the emitters~\cite{fischer2021controlled,tran2019suppression}.
Not only does the precise nature of these defect complexes remain under scrutiny~\cite{fischer2021controlled,mendelson2021identifying,li2021c_2c_n}, but also their quantum optical properties remain largely unexplored, particularly in regimes of strong light-matter coupling~\cite{vogl2019compact, haussler2021tunable}. 
Crucially, the optical properties of condensed matter systems cannot be understood through the standard theory of quantum optics, since strong system-environment interactions lead to a breakdown of the underpinning peturbative methods~\cite{Denning20phonons,Nazir16review}. 
Of particular relevance to the optical properties of defect complexes in 2DM are strong electron-phonon interactions. Specifically, defect emission spectra typically show sharp and well resolved phonon sidebands~\cite{vuong2016phonon,Feldman19phonon,Khatri19sideband,Grosso20lowtemp}, a hallmark of vibronic state formation and long lived correlations between electronic and vibrational degrees of freedom.
It is therefore crucial that any theory describing the dynamical or optical behaviour of a 2DM defect complex, accurately accounts for electron-phonon interactions.

To this end, we study the cavity quantum electrodynamics (cQED) of defects in 2DM~\cite{vogl2019compact,haussler2021tunable},  
focusing on two colour centres in hBN proposed as single-photon emitters, the \cbvn~\cite{sajid2020vncb,fischer2021controlled} and \ctcn~defect complexes~\cite{jara2021first,li2022carbon}.
We accurately account for electron-phonon interactions by combining \emph{ab initio} calculations of the defects, 
with a fully quantum mechanical and numerically exact description of the dynamics of the defect states interacting with a single mode optical cavity. 
The former provides an atomistic description of the phonon modes of the material system. 
The latter employs the time evolved matrix product operator (TEMPO) algorithm~\cite{strathearn2018efficient}, where non-Markovian influences of the phonon environment are encoded within a tensor network. When combined with tensor compression methods, TEMPO provides an efficient approach for calculating the reduced dynamics of the system~\cite{Jorgensen19causal,Gribben20structure,Gribben22exact}, which we extend to extract the linear response spectrum of the defect-cavity system.
This hybrid approach is inspired by \cite{svendsen2021combining} and allows us to maintain the accuracy and insight of first-principles calculations, while enabling coherent dynamics to be captured in previously inaccessible regimes of light-matter coupling, information that is typically inaccessible to first-principle methods.

A central finding of this work is that electron-phonon interactions in the studied defect complexes lead to highly non-Markovian dynamics, even in regimes of moderate light-matter coupling strengths, which are accessible to state-of-the-art experiments. 
{We attribute this behaviour to coherent coupling between the electronic states of the defect and high-quality ($Q$) factor phonon modes, leading to long-lived correlations between the system and its phonon environment.} 
This provides a mechanism to optically manipulate mechanical modes of a condensed matter system that would otherwise not directly couple to light.
Furthermore, from linear response spectra, we see evidence of hybridisation between cavity-defect polaritons and high-$Q$ phonon modes when entering regimes of strong light-matter coupling. 
These states bare resemblance to recently predicted exciton-photon-phonon hybridisation in hBN~\cite{Latini21phonoritons}, however in this instance they occur in a continuum of modes coupled to a defect state, and therefore will inherit non-linear features from the localised electronic states.

\begin{figure}
    \centering
    \includegraphics[width=\columnwidth]{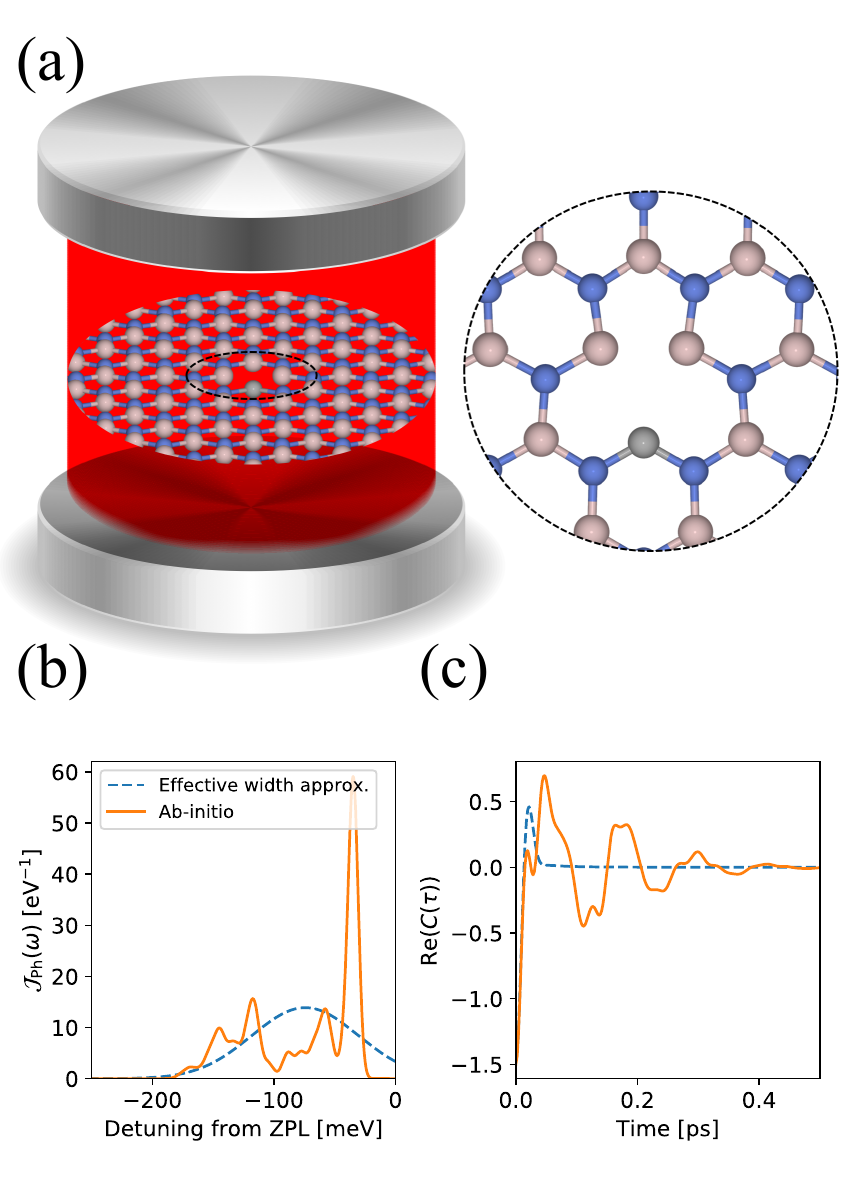}
    \caption{(a) A schematic figure of the \cbvn defect in hBN interacting with a single quantised field mode. This mode could be plasmonic or photonic in nature. 
    (b) A plot of the \emph{ab initio} (solid) and effective width (dashed) spectral densities. Using the atomistic simulation, we ascribe the dominant peak at $\sim125$~meV to a phonon breathing mode. 
    (c) shows the bath correlation function for the \emph{ab intio} (solid) and effective width (dashed) spectral densities. The sharp peaks of the \emph{ab initio} spectral density leads to long lived oscillations in the correlation function, which are washed out when an effective width approximation is used. 
    }
    \label{fig:fig1}
\end{figure}
For simplicity we focus on the \cbvn defect complex in the manuscript, leaving the discussion of C$_2$C$_N$ to the supplementary information (SI).
{The electronic structure of the \cbvn defect consists of two separate manifolds~\cite{Sajid18defect}: the first consists of the $(2)^1$A$_1$ and $(1)^1B_1$ excited states, and a $(1)^1A_1$ singlet ground state; the second contains the triplet states $(1)^3\mathrm{B}_1$ and $(2)^3\mathrm{B}_1$~\cite{li2021c_2c_n}, separated in energy by $\hbar\omega_e=2$~eV~\cite{Sajid18defect}. 
By assuming the inter-system crossing due to spin–orbit
coupling occurs on a much longer timescale ($10-100$~ns) than cavity-enhanced optical transitions ($<1$~ns), we restrict our attention to the triplet state manifold, which reduces to a two level system with ground and exited states, $\ket{g}$ and $\ket{e}$ respectively~\footnote{This analysis is repeated in the supplementary information for the C$_2$C$_N$ defect.}.}
The electronic transition interacts with a single mode optical cavity as shown schematically in Fig.~\ref{fig:fig1}a, with a cavity resonance $\hbar\Omega_c$ and described by a Jaynes-Cummings type interaction:
\begin{equation}
H_\mathrm{S} = \hbar\omega_e\sigma^\dagger\sigma + \hbar g(\sigma^\dagger a + \sigma a^\dagger) + \hbar\Omega_c a^\dagger a,
\label{eq:eq1}
\end{equation}
\noindent where $\sigma = \ket{g}\!\bra{e}$ is the transition dipole operator, $a$ ($a^\dagger$) is the annihilation (creation) operator of the cavity mode, and $g$ is the light-matter coupling strength.  
Here we limit ourselves to the single excitation subspace, spanned by the basis $\{\ket{g,0}, \ket{e,0},\ket{g,1}\}$. 
The composite cavity-emitter system interacts with both vibrational and electromagnetic environments, such that the global Hamiltonian is written as $H = H_\mathrm{S} + H_I^\mathrm{EM} + H_I^\mathrm{Ph}+ H_\mathrm{B}^\mathrm{EM} + H_\mathrm{B}^\mathrm{Ph}$.

The emitter-cavity system couples to an external electromagnetic environment either through direct emission from the defect into free space, or leakage from the cavity mode. In the rotating wave approximation the interaction to the electromagnetic field takes the form
$
    H_I^\mathrm{EM} = \sum_j \sum_\mathbf{l} (\hbar f_{j,\mathbf{l}} \hat{A}_j^\dagger c_{j,\mathbf{l}} + \mathrm{h.c.}),
$
where $\hat{A}_1=\sigma$ and $\hat{A}_2= a$, and $c_{j, \mathbf{l}}$ is the annihilation operator for the $\mathbf{l}^\mathrm{th}$~mode of the $j^\mathrm{th}$ electromagnetic environment. 
In the limit of weak coupling to these fields, we can make a flat spectrum approximation, such that $f_{1,\mathbf{1}} = \sqrt{2\pi\Gamma}$ and $f_{2,\mathbf{1}} = \sqrt{2\pi\kappa}$~\cite{carmichael1999statistical,maguire2019environmental}.
Here $\Gamma$ is the spontaneous emission rate from the TLE into free space, and $\kappa$ is the cavity leakage rate.
Using these approximations and tracing over the external EM fields, we can describe the optical contribution to the system evolution through the superoperator $\mathcal{V}_{t} = \exp(t~\mathcal{L}_\mathrm{S})$, where $\mathcal{L}_\mathrm{S}\rho = -i[H_\mathrm{S}, \rho] + \Gamma L_{\sigma}\rho + \kappa L_a\rho$, and we have introduced the Lindblad dissipators $L_o\rho = o\rho o^\dagger - \{o^\dagger o, \rho\}/2$.
For a full derivation and discussion of these approximations please refer to the SI.

We take a linear electron-phonon interaction~\cite{mahan2013many} of the form
$
    H_\mathrm{I}^\mathrm{Ph} = \sigma^\dagger\sigma\sum_{\mathbf{k}}\hbar g_\mathbf{k}(b_\mathbf{k}^\dagger + b_{-\mathbf{k}}), 
$
where $b_\mathbf{k}$ is the annihilation operator for a phonon mode with wave vector $\mathbf{k}$, with the corresponding free field evolution $H_\mathrm{B}^\mathrm{Ph} = \sum_\mathbf{k}\hbar\omega_\mathbf{k} b^\dagger_\mathbf{k} b_\mathbf{k}$, and $g_\mathbf{k}$ is the electron-phonon coupling strength of the $\mathbf{k}^\mathrm{th}$ phonon mode. 
The electron-phonon coupling can be fully characterised in terms of the spectral density $\mathcal{J}_\mathrm{Ph}(\omega)=\sum_\mathbf{k} S_\mathbf{k}\delta(\omega-\omega_\mathbf{k})$, where $S_\mathbf{k} = \omega_\mathbf{k}^{-2}\vert{}g_\mathbf{k}\vert^2$ is the  partial Huang-Rhys factor of mode $\mathbf{k}$. The total Huang-Rhys parameter can then be recovered by integrating the spectral density through $S_\mathrm{Tot}= \int_0^\infty d\omega \mathcal{J}_\mathrm{Ph}(\omega)$.

To determine the partial Huang-Rhys factors, and therefore the spectral density, we treat the electron-phonon coupling from first principles, using density functional theory (DFT) to calculate the normal modes of the ground- and excited states of the hBN lattice with the considered defect complex. 
Importantly, the normal modes of the ground- and excited states of the system can be very different and can even be of different types. To properly account for this, we employ the method outlined by Duschinsky \cite{borrelli2013franck} to calculate the partial Huang-Rhys factors. For completeness we also checked the generating function approach~\cite{alkauskas2014first} and found almost identical results, meaning that the modes in the ground and excited states for the system studied in the present work are in fact very similar.
The DFT calculations of the phonons were performed for periodically repeated defect complexes in hBN monolayers using the Vienna Ab Initio Simulation Package (VASP) \cite{kresse1993ab}. In order to avoid interactions between the periodic copies of the defects, we used a 9x9x1 supercell for the calculations, which we relax on a 3x3x1 K-point grid, to a force convergence of $10^{-3}$ eV Å$^{-1}$ using a plane wave cut-off of 700 eV. All calculations were performed with the HSE06 exchange correlation functional, which is essential for an accurate description of the electron-phonon coupling \cite{heyd2004efficient}. 
The normal modes and the dynamical matrix was calculated at the $\Gamma$-point. %
Note that we relax the defect in the in-plane constrained C$_2V$ configuration.

\begin{figure*}[htb]
    \centering
    \includegraphics[width=2\columnwidth]{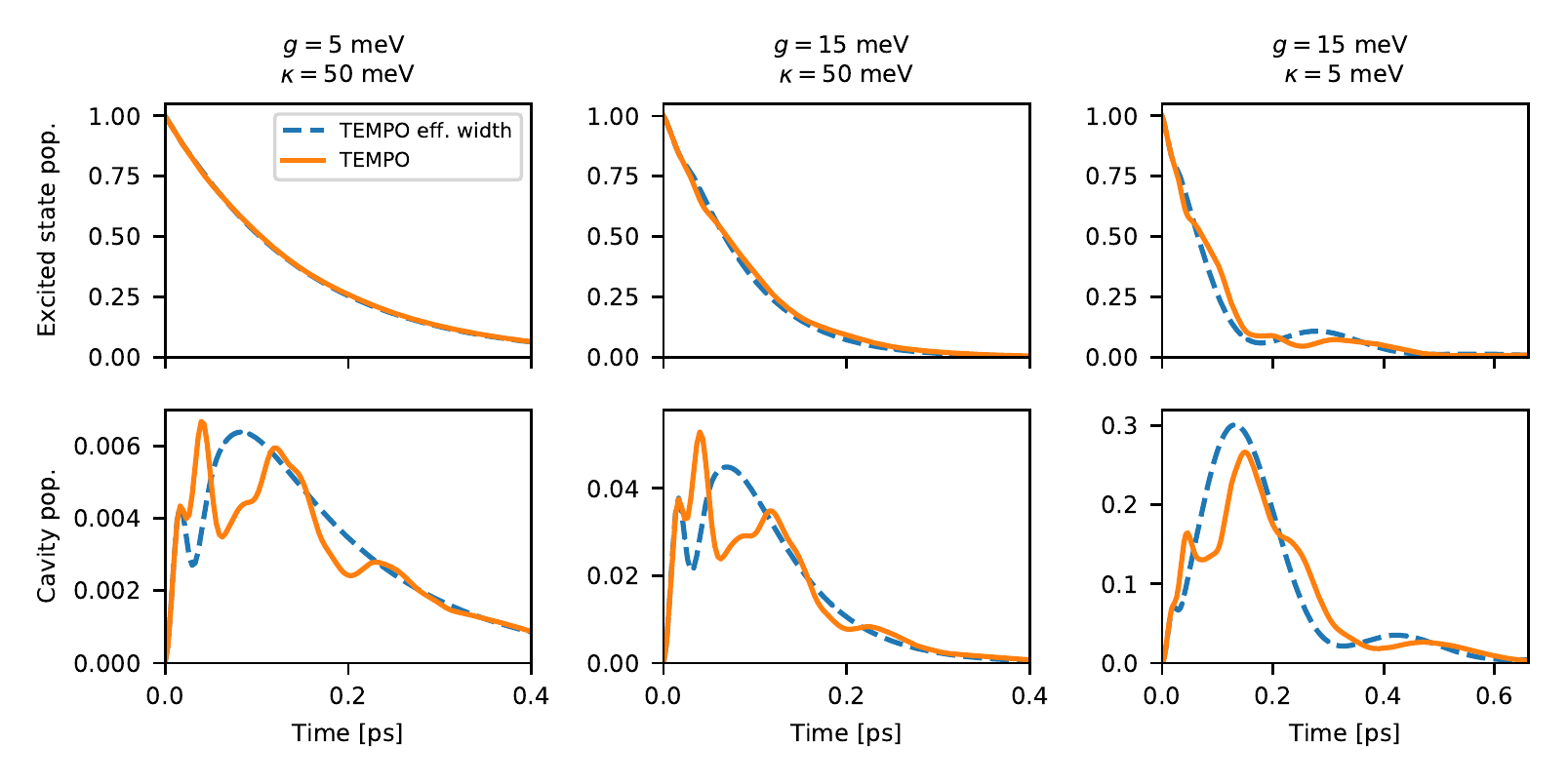}
    \caption{Time dependent emitter population (top) and cavity occupation (bottom) for different coupling strengths and cavity widths. In contrast to the effective width spectral density (dashed), the \emph{ab initio} spectral density (solid orange) predicts significant oscillations in the cavity occupation over all parameter regimes. Parameters used are $T=4$~K, $\Gamma=4$~meV, and $\hbar\Omega_c =\hbar \omega_e - \hbar\lambda$ where $\lambda=\int_0^\infty d\omega \omega^{-2}\mathcal{J}_\mathrm{Ph}(\omega)$ is the reorganisation energy. The step size used to obtain convergence was $dt=3.2$ fs, with SVD cut-off $\epsilon_\mathrm{C}=10^{-6}$.}
    \label{fig:fig2}
\end{figure*}

{The resulting spectral density is shown in Fig.~\ref{fig:fig1}(b), where we see multiple sharp peaks present. 
These peaks correspond to high-$Q$ phonon modes present in hBN; {notably from the atomistic simulations, we can assign the dominant contributions to the peak $\sim125~$meV from de-localized defect breathing modes, in which the atoms surrounding the defect oscillate along the dipole direction of the defect.}

To understand the influence of these modes will have on the cavity-defect system, we can calculate the bath correlation function $C(\tau) = \int_0^\infty d\omega \mathcal{J}_\mathrm{Ph}(\omega)\left(\cos(\omega\tau)\coth(\omega/2k_\mathrm{B}T)-i\sin(\omega\tau)\right)$, shown in Fig.~\ref{fig:fig1}(c), where $T$ is the temperature, $k_\mathrm{B}$ is the Boltzmann constant, and we have assumed the phonon bath to be initially in thermal equilibrium.
This correlation function encodes the timescales over which memory effects last between the system and environment~\cite{breuer2002theory}, for a full derivation we refer the reader to the SI.
The sharp peaks in the \emph{ab intio} spectral density lead to long lived oscillation in the correlation function, which can be attributed to phonon modes with large $Q$-factors. To highlight the importance of this structure we can replace the spectral density with a broadened function with the same total Huang-Rhys parameter, as shown by the dashed curve in Fig.~\ref{fig:fig1}(b). The corresponding correlation function, shown in Fig.~\ref{fig:fig1}(c), has only small oscillations which are rapidly damped. }

To study the time evolution of the composite cavity-emitter system, we employ the time-evolving matrix product operator (TEMPO) algorithm initially developed by Strathearn~\emph{et al}~\cite{strathearn2018efficient}.
TEMPO is a powerful method for the study of open quantum systems in strong coupling regimes~\cite{Jorgensen19causal,Gribben20structure}, and has been applied to study quantum thermodynamics in the strong coupling regime~\cite{Popovic21heat}, as well as non-additive phenomena in quantum optics~\cite{Gribben22exact}, and optimal control~\cite{Fux21efficient}. 
The starting point for the TEMPO formalism, alongside other numerically exact path integral methods~\cite{makri1995tensorI,makri1995tensorII,Cygorek22ace}, is the Trotter decomposition~\cite{trotter1959product}, where for a sufficiently small time increment $\delta t$, the propagator for the open system can be factorised such that $\mathcal{U}_{\delta t} = e^{(\mathcal{L}_S + \mathcal{L}_B)\delta t}\approx \mathcal{V}_{\delta t}^{1/2} \mathcal{W}_{\delta t} \mathcal{V}_{\delta t}^{1/2} + \mathcal{O}(\delta t^3)$. 
The superoperator $\mathcal{V}_{\delta t}$ is as defined above, and captures the evolution of the system and dissipation through the external electromagnetic fields.
The interaction between the system and phonon environment is captured through $\mathcal{W}_{\delta t} = \exp(\delta t \mathcal{L}_\mathrm{B})$, and is given by $\mathcal{L}_\mathrm{B}\rho = -i\left[H_\mathrm{I}^\mathrm{Ph} + H_\mathrm{B}^\mathrm{Ph},\rho\right]$. 

This partitioning allows one to construct a discrete-time influence functional of Feynman-Vernon type~\cite{makri1995tensorI,makri1995tensorII}, which captures the influence of the environment to all orders in the interaction strength. 
The reduced state after $k$-time-steps can be expressed as: 
\begin{equation}
    \rho^{\alpha_k} = \sum_{\vec\alpha, \vec\beta} \rho^{\alpha_0}\prod_{j=1}^k\mathcal{V}_{\beta_{j}}^{\alpha_j}\mathcal{V}^{\beta_{j}}_{\alpha_{j-1}}\mathcal{F}_{\beta_{k}\cdots\beta_1},
    \label{eq:reduced}
\end{equation}
where we have introduced the compound indices $\alpha_k = (s_k, r_k)$ and $\beta_k = (t_k, u_k)$, yielding the density matrix elements $\rho^{\alpha_k} = \bra{r_k}\rho\ket{s_k}$, and the system superoperator is given by $\mathcal{V}^{\alpha_k}_{\beta_k} = \bra{r_k}\mathcal{V}_{\delta t}[\ket{t_k}\bra{u_k}]\ket{s_k}$.
The influence of the phonon environment is captured by the influence tensor $\mathcal{F}_{\beta_k\cdots\beta_1}=\tr_E(\mathcal{W}^{\beta_k}\dots\mathcal{W}^{\beta_1}[\tau_B])$, where $\mathcal{W}^{\beta_k} =\bra{t_k}\mathcal{W}_{\delta t}[\ket{t_k}\bra{u_k}]\ket{u_k} $. By taking the initial state of the environment to be in thermal equilibrium $\tau_B = \exp(-\sum_k \nu_k b_k^\dagger b_k/k_BT)/\tr[ \exp(-\sum_k \nu_k b_k^\dagger b_k/k_BT)]$, the trace over the environmental degrees of freedom can be done analytically~\cite{makri1995tensorI,makri1995tensorII}.

Crucially, the influence tensor scales exponentially in the number of time-steps taken~\cite{makri1995tensorI, makri1995tensorII}; while applying a finite time memory approximation can reduce the computational cost of propagating the reduced state of the system out to long times~\cite{makri1995tensorI,makri1995tensorII,Strathearn17quapi}, it limits one to scenarios where key dynamics occur on short timescales. 
A key insight of Strathearn~\emph{et al}~\cite{strathearn2018efficient} was that the influence tensor may be represented in matrix product operator (MPO) form. This allows one to encode the exponentially growing tensor as a tensor network, and apply tensor compression~\cite{orus14intro} to reduce the rank of the elements in the network, thereby circumventing exponential scaling. 
The reduced state of the electronic system is then calculated by contracting the network down after each time-step. The convergence of the TEMPO algorithm is sensitive to taking a sufficiently small timestep $\delta t$, and the degree to which the tensors are compressed. 
Further details of TEMPO and its convergence properties are discussed in the SI.

We now consider the dynamics of the cavity-defect system initialised in the state $\rho(0) = \ket{e,0}\!\bra{e,0}$.
Fig.~\ref{fig:fig2} shows the time evolution of the excited state population and the cavity mode occupation for different cavity coupling parameters. 
We find that across all parameter regimes TEMPO predicts complex oscillations in the cavity occupation.
We attribute these oscillations to high-$Q$ phonon modes in the environment, which lead to long-lived oscillations in the bath correlation function shown in Fig.~\ref{fig:fig1}c, indeed these oscillations follow closely those observed in the bath correlation function. 
We confirm this by repeating the TEMPO calculations with the effective width spectral density shown in Fig.~\ref{fig:fig1}b. 
Since this spectral density has the same total Huang-Rhys parameter, na\"ively we might expect similar dynamics to emerge.
However, the resulting dynamics in Fig.~\ref{fig:fig2} show no oscillations at weak light-matter coupling, suggesting it is indeed the structure in the spectral density that has the dominant contribution to non-Markovian behaviour.
Interestingly, phonon induced oscillations are present in the cavity dynamics across all parameter regimes studied, but only emerge in the emitter dynamics in the strong light-matter coupling regime. 
This is a consequence of phonon processes inducing fluctuations to the defects' excited state energy; the cavity is sensitive to these fluctuations through the dipole coupling with the emitter, leading to the observed oscillations in the occupation. 
However, these effects do become increasingly visible in the emitter dynamics with increasing coupling strength, with the onset of vacuum Rabi oscillations.

We now turn our attention to the linear spectrum of the cavity-defect system. 
The linear response of an electronic system coupled to a quantized mode can be formulated in two ways, by an external field coupling to the electronic degrees of freedom, or by an external current pumping the cavity mode~\cite{ruggenthaler2018quantum,flick2019light}. 
In both cases the driving can be included as a perturbation at the Hamiltonian level, such that $H^\prime(t)=H + E(t)\cdot\mu$. Here the system transition operator can take on two values $\mu=\{\sigma^\dagger + \sigma, a^\dagger + a\}$. The first is the dipole operator of the defect, describing the scattering of light directly off the two level transition, and forms the basis of standard linear response theory~\cite{mukamel1999principles}. 
The second is the quadrature operator of the cavity field, which induces a static polarisation of the cavity mode and excites real photons into the field~\cite{flick2019light}. $E(t)$ then corresponds to the external field or the external current respectively.

\begin{figure}
    \centering
    \includegraphics[width=\columnwidth]{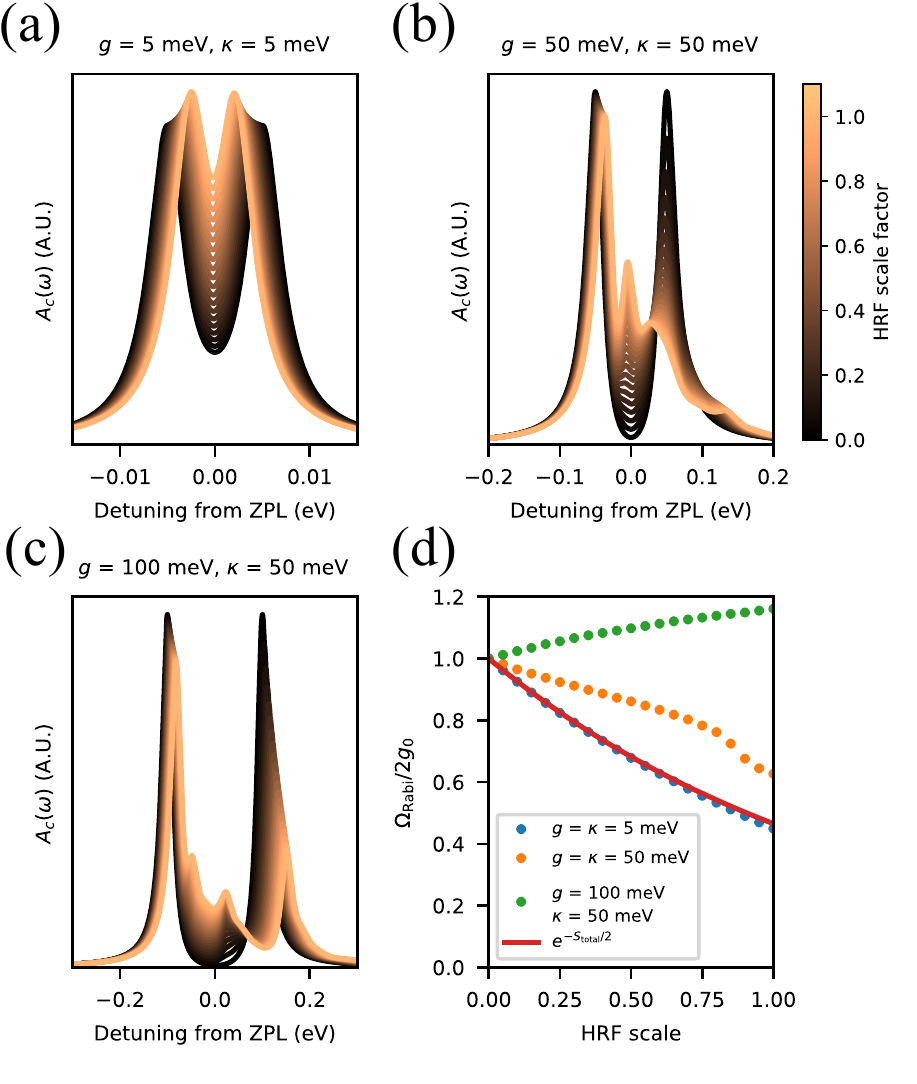}
    \caption{(a-c) the linear absorption spectra when probing the cavity mode, for various values of the scaling parameter $\alpha_\mathrm{HRF}$. Here structure in the absorption spectra becomes increasingly important for large coupling, and broad cavity widths. (d) shows the change in peak position as a function of $\alpha_\mathrm{HRF}$ (points) for the cavity parameters studied in (a-c). All other parameters are the same as Fig.~\ref{fig:fig2}}
    \label{fig:fig3}
\end{figure}
Treating the driving as weak, we can extract the linear response spectrum using density matrix perturbation theory~\cite{mukamel1999principles}. Taking the semi-impulsive limit such that $E(t)\propto e^{i\omega_D t}\delta(t)$, where $\omega_D$ is the driving frequency, we obtain the absorption spectrum:
\begin{equation}
    A(\omega)= 2\mathrm{Re}\left[\int_0^\infty dt e^{i\omega t}S^{(1)}(t)\right],
\end{equation}
where we have introduced the first-order response function $S^{(1)}(t) = \tr(\mu(t)[\mu(0),\chi(-\infty)])$. The global equilibrium density operator is given by $\chi(-\infty) = \ket{g,0}\!\bra{g,0}\otimes\tau_B\otimes\ket{\{0\}}\!\bra{\{0\}}$, where $\ket{\{0\}}$ denotes the vacuum state of the electromagnetic field. 
The above expressions allow us to extend TEMPO to calculate the first-order response function by propagating the initial system state $\rho(0)=\mu(0)\ket{g,0}\!\bra{g,0}$. 

Figs.~\ref{fig:fig3}(a-c) show the absorption spectra of the cavity mode when driven by an external current for various cavity parameters. To understand the role of phonons in these spectra, we artificially include a scaling parameter to spectral density $\mathcal{J}_\mathrm{Ph}(\omega)\rightarrow\alpha_\mathrm{HRF} \mathcal{J}_\mathrm{Ph}(\omega)$, where $\alpha_\mathrm{HRF}\in[0,1]$, equivalent to scaling the total Huang-Rhys parameter.
Fig.~\ref{fig:fig3}a shows absorption spectrum at the onset of the strong coupling regime, where $g=\kappa$. 
In the absence of phonons ($\alpha_\mathrm{HRF}=0$) we see a clear Rabi splitting, with the separation of the peaks determined by the the light-matter coupling $g$. As $\alpha_\mathrm{HRF}$ increases, the separation of these peaks is reduced. We can understand this reduction by appealing to the polaron formalism commonly used to study the behaviour semi-conductor quantum dots~\cite{Nazir16review,iles2017phonon}: here, the light-matter coupling strength is reduced by the Frank-Condon factor $\mathscr{F} = \exp(-\alpha_\mathrm{HRF} S_\mathrm{Tot}/2)$, which accounts for the geometric difference of the phonon modes associated to the emitter ground and excited states. 
Fig.~\ref{fig:fig3}d compares the peak separation as a function of the scaling parameter $\alpha_\mathrm{HRF}$ for the three parameter regimes. For the parameters associated with Fig.~3a, the peak separation follows the Frank-Condon factor $\mathscr{F}$ (solid curve).  

In regimes of stronger light-matter coupling show in Fig.~\ref{fig:fig3}b and c, a more complex picture emerges: significant structure becomes apparent in the spectra as $\alpha_\mathrm{HRF}$ increases. 
Of particular interest is the departure from the well understood polaronic physics seen in Fig.~\ref{fig:fig3}a, which is highlighted in Fig.~\ref{fig:fig3}d. 
Here we see that at stronger coupling regimes ($g=50$~meV), the renormalisation no longer follows the Frank-Condon factor, and in regime of strong light-matter coupling ($g=100$~meV), we in-fact see the splitting increase with $\alpha_\mathrm{HRF}$. 
We attribute this behaviour to a hybridisation of the light-matter polariton and high-$Q$ phonon modes. 
The resultant state is a tri-partite quasi-particle with characteristics of light, matter, and vibrations.
This can be seen most clearly in the upper-polariton of Fig.~\ref{fig:fig3}c, where an additional splitting emerges when $\alpha_\mathrm{HRF}\rightarrow1$. 
This interpretation is supported by the C$_2$C$_N$ calculations shown in the SI; this defect complex has a spectral density with little low energy structure, such that at $g=50$~meV no hybridisation occurs, and the renormalisation of the Rabi frequency follows closely the Frank-Condon factor $\mathscr{F}$. It is only at higher light-matter coupling strengths ($g=100$~meV) for C$_2$C$_N$, when the polariton splitting approaches resonance with a high-Q phonon mode that we observe a departure from Frank Condon physics, heralding hybridised polariton-polaron states. 

\emph{Conclusion---} In this letter we have combined atomistic simulations of a defect complex in hBN with TEMPO, a numerically exact and fully quantum mechanical simulation method. 
Our hybrid approach allows us to study realistic emitters beyond phenomenological and approximate treatments~\cite{Feldman19phonon}, providing a complete description of electron-phonon interactions in condensed matter single photon emitters in optical or plasmonic cavities, with direct relevance to on-going experiments. 
Furthermore, by considering the cavity quantum electrodynamics of a defect across the weak and strong light-matter coupling regimes,
we have shown that strong coupling to high-Q vibrational modes play a significant role in determining the dynamics of the cavity-defect system, even in the weak light-matter coupling regime. 
At strong light-matter coupling, the absorption spectrum show clear signatures of hybridisation between the light-matter polaritons and phonon modes inherent to hBN.

The method we present here is general, and not restricted to the material system or specific defect complexes studied here, with potential application to organic polaritons~\cite{del2018tensor}.
It is worth noting that we have restricted ourselves to low temperatures in the above discussion; at elevated temperatures phonon processes beyond linear electron-phonon coupling become important through mechanisms such as the Jahn-Teller effect~\cite{white2021phonon}.

\emph{Acknowledgments---}
 We thank Mortiz Fischer and Emil V. Denning for insightful conversations.
The work is supported by the BIOMAG project funded by the Novo Nordisk Foundation (grant no. NNF21OC0066526)
We also
acknowledge funding from the European Research Council (ERC) under the European Union’s
Horizon 2020 research and innovation program Grant No. 773122 (LIMA). K. S. T. is a Villum
Investigator supported by Villum Foundation (grant no. 37789).
 This work was supported by the Danish National Research Foundation through NanoPhoton - Center
for Nanophotonics, Grant No. DNRF147 and Center for Nanostructured Graphene, Grant No. DNRF103. 
NS acknowledges support from the Villum Foundation through grant no.
00028233 and from the Independent Research Fund Denmark - Natural Sciences (Project no. 0135-00403B).
 J.I.-S. acknowledges support from the Royal Commission for the Exhibition of 1851.

\bibliography{references}

\begin{thebibliography}{57}%
\makeatletter
\providecommand \@ifxundefined [1]{%
 \@ifx{#1\undefined}
}%
\providecommand \@ifnum [1]{%
 \ifnum #1\expandafter \@firstoftwo
 \else \expandafter \@secondoftwo
 \fi
}%
\providecommand \@ifx [1]{%
 \ifx #1\expandafter \@firstoftwo
 \else \expandafter \@secondoftwo
 \fi
}%
\providecommand \natexlab [1]{#1}%
\providecommand \enquote  [1]{``#1''}%
\providecommand \bibnamefont  [1]{#1}%
\providecommand \bibfnamefont [1]{#1}%
\providecommand \citenamefont [1]{#1}%
\providecommand \href@noop [0]{\@secondoftwo}%
\providecommand \href [0]{\begingroup \@sanitize@url \@href}%
\providecommand \@href[1]{\@@startlink{#1}\@@href}%
\providecommand \@@href[1]{\endgroup#1\@@endlink}%
\providecommand \@sanitize@url [0]{\catcode `\\12\catcode `\$12\catcode
  `\&12\catcode `\#12\catcode `\^12\catcode `\_12\catcode `\%12\relax}%
\providecommand \@@startlink[1]{}%
\providecommand \@@endlink[0]{}%
\providecommand \url  [0]{\begingroup\@sanitize@url \@url }%
\providecommand \@url [1]{\endgroup\@href {#1}{\urlprefix }}%
\providecommand \urlprefix  [0]{URL }%
\providecommand \Eprint [0]{\href }%
\providecommand \doibase [0]{https://doi.org/}%
\providecommand \selectlanguage [0]{\@gobble}%
\providecommand \bibinfo  [0]{\@secondoftwo}%
\providecommand \bibfield  [0]{\@secondoftwo}%
\providecommand \translation [1]{[#1]}%
\providecommand \BibitemOpen [0]{}%
\providecommand \bibitemStop [0]{}%
\providecommand \bibitemNoStop [0]{.\EOS\space}%
\providecommand \EOS [0]{\spacefactor3000\relax}%
\providecommand \BibitemShut  [1]{\csname bibitem#1\endcsname}%
\let\auto@bib@innerbib\@empty
\bibitem [{\citenamefont {Chakraborty}\ \emph {et~al.}(2015)\citenamefont
  {Chakraborty}, \citenamefont {Kinnischtzke}, \citenamefont {Goodfellow},
  \citenamefont {Beams},\ and\ \citenamefont
  {Vamivakas}}]{Chakraborty2015wse2}%
  \BibitemOpen
  \bibfield  {author} {\bibinfo {author} {\bibfnamefont {C.}~\bibnamefont
  {Chakraborty}}, \bibinfo {author} {\bibfnamefont {L.}~\bibnamefont
  {Kinnischtzke}}, \bibinfo {author} {\bibfnamefont {K.~M.}\ \bibnamefont
  {Goodfellow}}, \bibinfo {author} {\bibfnamefont {R.}~\bibnamefont {Beams}},\
  and\ \bibinfo {author} {\bibfnamefont {A.~N.}\ \bibnamefont {Vamivakas}},\
  }\bibfield  {title} {\bibinfo {title} {Voltage-controlled quantum light from
  an atomically thin semiconductor},\ }\href
  {https://doi.org/10.1038/nnano.2015.79} {\bibfield  {journal} {\bibinfo
  {journal} {Nature Nanotechnology}\ }\textbf {\bibinfo {volume} {10}},\
  \bibinfo {pages} {507} (\bibinfo {year} {2015})}\BibitemShut {NoStop}%
\bibitem [{\citenamefont {He}\ \emph {et~al.}(2015)\citenamefont {He},
  \citenamefont {Clark}, \citenamefont {Schaibley}, \citenamefont {He},
  \citenamefont {Chen}, \citenamefont {Wei}, \citenamefont {Ding},
  \citenamefont {Zhang}, \citenamefont {Yao}, \citenamefont {Xu}, \citenamefont
  {Lu},\ and\ \citenamefont {Pan}}]{He2015wse2}%
  \BibitemOpen
  \bibfield  {author} {\bibinfo {author} {\bibfnamefont {Y.-M.}\ \bibnamefont
  {He}}, \bibinfo {author} {\bibfnamefont {G.}~\bibnamefont {Clark}}, \bibinfo
  {author} {\bibfnamefont {J.~R.}\ \bibnamefont {Schaibley}}, \bibinfo {author}
  {\bibfnamefont {Y.}~\bibnamefont {He}}, \bibinfo {author} {\bibfnamefont
  {M.-C.}\ \bibnamefont {Chen}}, \bibinfo {author} {\bibfnamefont {Y.-J.}\
  \bibnamefont {Wei}}, \bibinfo {author} {\bibfnamefont {X.}~\bibnamefont
  {Ding}}, \bibinfo {author} {\bibfnamefont {Q.}~\bibnamefont {Zhang}},
  \bibinfo {author} {\bibfnamefont {W.}~\bibnamefont {Yao}}, \bibinfo {author}
  {\bibfnamefont {X.}~\bibnamefont {Xu}}, \bibinfo {author} {\bibfnamefont
  {C.-Y.}\ \bibnamefont {Lu}},\ and\ \bibinfo {author} {\bibfnamefont {J.-W.}\
  \bibnamefont {Pan}},\ }\bibfield  {title} {\bibinfo {title} {Single quantum
  emitters in monolayer semiconductors},\ }\href
  {https://doi.org/10.1038/nnano.2015.75} {\bibfield  {journal} {\bibinfo
  {journal} {Nature Nanotechnology}\ }\textbf {\bibinfo {volume} {10}},\
  \bibinfo {pages} {497} (\bibinfo {year} {2015})}\BibitemShut {NoStop}%
\bibitem [{\citenamefont {Koperski}\ \emph {et~al.}(2015)\citenamefont
  {Koperski}, \citenamefont {Nogajewski}, \citenamefont {Arora}, \citenamefont
  {Cherkez}, \citenamefont {Mallet}, \citenamefont {Veuillen}, \citenamefont
  {Marcus}, \citenamefont {Kossacki},\ and\ \citenamefont
  {Potemski}}]{Koperski2015wse2}%
  \BibitemOpen
  \bibfield  {author} {\bibinfo {author} {\bibfnamefont {M.}~\bibnamefont
  {Koperski}}, \bibinfo {author} {\bibfnamefont {K.}~\bibnamefont
  {Nogajewski}}, \bibinfo {author} {\bibfnamefont {A.}~\bibnamefont {Arora}},
  \bibinfo {author} {\bibfnamefont {V.}~\bibnamefont {Cherkez}}, \bibinfo
  {author} {\bibfnamefont {P.}~\bibnamefont {Mallet}}, \bibinfo {author}
  {\bibfnamefont {J.-Y.}\ \bibnamefont {Veuillen}}, \bibinfo {author}
  {\bibfnamefont {J.}~\bibnamefont {Marcus}}, \bibinfo {author} {\bibfnamefont
  {P.}~\bibnamefont {Kossacki}},\ and\ \bibinfo {author} {\bibfnamefont
  {M.}~\bibnamefont {Potemski}},\ }\bibfield  {title} {\bibinfo {title} {Single
  photon emitters in exfoliated wse2 structures},\ }\href
  {https://doi.org/10.1038/nnano.2015.67} {\bibfield  {journal} {\bibinfo
  {journal} {Nature Nanotechnology}\ }\textbf {\bibinfo {volume} {10}},\
  \bibinfo {pages} {503} (\bibinfo {year} {2015})}\BibitemShut {NoStop}%
\bibitem [{\citenamefont {Srivastava}\ \emph {et~al.}(2015)\citenamefont
  {Srivastava}, \citenamefont {Sidler}, \citenamefont {Allain}, \citenamefont
  {Lembke}, \citenamefont {Kis},\ and\ \citenamefont
  {Imamoğlu}}]{Srivastava2015wse2}%
  \BibitemOpen
  \bibfield  {author} {\bibinfo {author} {\bibfnamefont {A.}~\bibnamefont
  {Srivastava}}, \bibinfo {author} {\bibfnamefont {M.}~\bibnamefont {Sidler}},
  \bibinfo {author} {\bibfnamefont {A.~V.}\ \bibnamefont {Allain}}, \bibinfo
  {author} {\bibfnamefont {D.~S.}\ \bibnamefont {Lembke}}, \bibinfo {author}
  {\bibfnamefont {A.}~\bibnamefont {Kis}},\ and\ \bibinfo {author}
  {\bibfnamefont {A.}~\bibnamefont {Imamoğlu}},\ }\bibfield  {title} {\bibinfo
  {title} {Optically active quantum dots in monolayer $\mathrm{WSe}_2$},\
  }\href {https://doi.org/10.1038/nnano.2015.60} {\bibfield  {journal}
  {\bibinfo  {journal} {Nature Nanotechnology}\ }\textbf {\bibinfo {volume}
  {10}},\ \bibinfo {pages} {491} (\bibinfo {year} {2015})}\BibitemShut
  {NoStop}%
\bibitem [{\citenamefont {Klein}\ \emph {et~al.}(2019)\citenamefont {Klein},
  \citenamefont {Lorke}, \citenamefont {Florian}, \citenamefont {Sigger},
  \citenamefont {Sigl}, \citenamefont {Rey}, \citenamefont {Wierzbowski},
  \citenamefont {Cerne}, \citenamefont {M{\"u}ller}, \citenamefont
  {Mitterreiter} \emph {et~al.}}]{klein2019site}%
  \BibitemOpen
  \bibfield  {author} {\bibinfo {author} {\bibfnamefont {J.}~\bibnamefont
  {Klein}}, \bibinfo {author} {\bibfnamefont {M.}~\bibnamefont {Lorke}},
  \bibinfo {author} {\bibfnamefont {M.}~\bibnamefont {Florian}}, \bibinfo
  {author} {\bibfnamefont {F.}~\bibnamefont {Sigger}}, \bibinfo {author}
  {\bibfnamefont {L.}~\bibnamefont {Sigl}}, \bibinfo {author} {\bibfnamefont
  {S.}~\bibnamefont {Rey}}, \bibinfo {author} {\bibfnamefont {J.}~\bibnamefont
  {Wierzbowski}}, \bibinfo {author} {\bibfnamefont {J.}~\bibnamefont {Cerne}},
  \bibinfo {author} {\bibfnamefont {K.}~\bibnamefont {M{\"u}ller}}, \bibinfo
  {author} {\bibfnamefont {E.}~\bibnamefont {Mitterreiter}}, \emph {et~al.},\
  }\bibfield  {title} {\bibinfo {title} {Site-selectively generated photon
  emitters in monolayer mos2 via local helium ion irradiation},\ }\href@noop {}
  {\bibfield  {journal} {\bibinfo  {journal} {Nature communications}\ }\textbf
  {\bibinfo {volume} {10}},\ \bibinfo {pages} {1} (\bibinfo {year}
  {2019})}\BibitemShut {NoStop}%
\bibitem [{\citenamefont {Michaelis~de Vasconcellos}\ \emph
  {et~al.}(2022)\citenamefont {Michaelis~de Vasconcellos}, \citenamefont
  {Wigger}, \citenamefont {Wurstbauer}, \citenamefont {Holleitner},
  \citenamefont {Bratschitsch},\ and\ \citenamefont
  {Kuhn}}]{michaelis2022single}%
  \BibitemOpen
  \bibfield  {author} {\bibinfo {author} {\bibfnamefont {S.}~\bibnamefont
  {Michaelis~de Vasconcellos}}, \bibinfo {author} {\bibfnamefont
  {D.}~\bibnamefont {Wigger}}, \bibinfo {author} {\bibfnamefont
  {U.}~\bibnamefont {Wurstbauer}}, \bibinfo {author} {\bibfnamefont {A.~W.}\
  \bibnamefont {Holleitner}}, \bibinfo {author} {\bibfnamefont
  {R.}~\bibnamefont {Bratschitsch}},\ and\ \bibinfo {author} {\bibfnamefont
  {T.}~\bibnamefont {Kuhn}},\ }\bibfield  {title} {\bibinfo {title}
  {Single-photon emitters in layered van der waals materials},\ }\href@noop {}
  {\bibfield  {journal} {\bibinfo  {journal} {physica status solidi (b)}\ ,\
  \bibinfo {pages} {2100566}} (\bibinfo {year} {2022})}\BibitemShut {NoStop}%
\bibitem [{\citenamefont {Kianinia}\ \emph {et~al.}(2022)\citenamefont
  {Kianinia}, \citenamefont {Xu}, \citenamefont {Toth},\ and\ \citenamefont
  {Aharonovich}}]{kianinia2022quantum}%
  \BibitemOpen
  \bibfield  {author} {\bibinfo {author} {\bibfnamefont {M.}~\bibnamefont
  {Kianinia}}, \bibinfo {author} {\bibfnamefont {Z.-Q.}\ \bibnamefont {Xu}},
  \bibinfo {author} {\bibfnamefont {M.}~\bibnamefont {Toth}},\ and\ \bibinfo
  {author} {\bibfnamefont {I.}~\bibnamefont {Aharonovich}},\ }\bibfield
  {title} {\bibinfo {title} {Quantum emitters in 2d materials: Emitter
  engineering, photophysics, and integration in photonic nanostructures},\
  }\href@noop {} {\bibfield  {journal} {\bibinfo  {journal} {Applied Physics
  Reviews}\ }\textbf {\bibinfo {volume} {9}},\ \bibinfo {pages} {011306}
  (\bibinfo {year} {2022})}\BibitemShut {NoStop}%
\bibitem [{\citenamefont {Tran}\ \emph {et~al.}(2016)\citenamefont {Tran},
  \citenamefont {Bray}, \citenamefont {Ford}, \citenamefont {Toth},\ and\
  \citenamefont {Aharonovich}}]{Tran2016hbn}%
  \BibitemOpen
  \bibfield  {author} {\bibinfo {author} {\bibfnamefont {T.~T.}\ \bibnamefont
  {Tran}}, \bibinfo {author} {\bibfnamefont {K.}~\bibnamefont {Bray}}, \bibinfo
  {author} {\bibfnamefont {M.~J.}\ \bibnamefont {Ford}}, \bibinfo {author}
  {\bibfnamefont {M.}~\bibnamefont {Toth}},\ and\ \bibinfo {author}
  {\bibfnamefont {I.}~\bibnamefont {Aharonovich}},\ }\bibfield  {title}
  {\bibinfo {title} {Quantum emission from hexagonal boron nitride
  monolayers},\ }\href {https://doi.org/10.1038/nnano.2015.242} {\bibfield
  {journal} {\bibinfo  {journal} {Nature Nanotechnology}\ }\textbf {\bibinfo
  {volume} {11}},\ \bibinfo {pages} {37} (\bibinfo {year} {2016})}\BibitemShut
  {NoStop}%
\bibitem [{\citenamefont {Aharonovich}\ and\ \citenamefont
  {Toth}(2017)}]{aharonovich2017quantum}%
  \BibitemOpen
  \bibfield  {author} {\bibinfo {author} {\bibfnamefont {I.}~\bibnamefont
  {Aharonovich}}\ and\ \bibinfo {author} {\bibfnamefont {M.}~\bibnamefont
  {Toth}},\ }\bibfield  {title} {\bibinfo {title} {Quantum emitters in two
  dimensions},\ }\href@noop {} {\bibfield  {journal} {\bibinfo  {journal}
  {Science}\ }\textbf {\bibinfo {volume} {358}},\ \bibinfo {pages} {170}
  (\bibinfo {year} {2017})}\BibitemShut {NoStop}%
\bibitem [{\citenamefont {Fischer}\ \emph {et~al.}(2021)\citenamefont
  {Fischer}, \citenamefont {Caridad}, \citenamefont {Sajid}, \citenamefont
  {Ghaderzadeh}, \citenamefont {Ghorbani-Asl}, \citenamefont {Gammelgaard},
  \citenamefont {B{\o}ggild}, \citenamefont {Thygesen}, \citenamefont
  {Krasheninnikov}, \citenamefont {Xiao} \emph
  {et~al.}}]{fischer2021controlled}%
  \BibitemOpen
  \bibfield  {author} {\bibinfo {author} {\bibfnamefont {M.}~\bibnamefont
  {Fischer}}, \bibinfo {author} {\bibfnamefont {J.}~\bibnamefont {Caridad}},
  \bibinfo {author} {\bibfnamefont {A.}~\bibnamefont {Sajid}}, \bibinfo
  {author} {\bibfnamefont {S.}~\bibnamefont {Ghaderzadeh}}, \bibinfo {author}
  {\bibfnamefont {M.}~\bibnamefont {Ghorbani-Asl}}, \bibinfo {author}
  {\bibfnamefont {L.}~\bibnamefont {Gammelgaard}}, \bibinfo {author}
  {\bibfnamefont {P.}~\bibnamefont {B{\o}ggild}}, \bibinfo {author}
  {\bibfnamefont {K.~S.}\ \bibnamefont {Thygesen}}, \bibinfo {author}
  {\bibfnamefont {A.}~\bibnamefont {Krasheninnikov}}, \bibinfo {author}
  {\bibfnamefont {S.}~\bibnamefont {Xiao}}, \emph {et~al.},\ }\bibfield
  {title} {\bibinfo {title} {Controlled generation of luminescent centers in
  hexagonal boron nitride by irradiation engineering},\ }\href@noop {}
  {\bibfield  {journal} {\bibinfo  {journal} {Science Advances}\ }\textbf
  {\bibinfo {volume} {7}},\ \bibinfo {pages} {eabe7138} (\bibinfo {year}
  {2021})}\BibitemShut {NoStop}%
\bibitem [{\citenamefont {Mendelson}\ \emph {et~al.}(2021)\citenamefont
  {Mendelson}, \citenamefont {Chugh}, \citenamefont {Reimers}, \citenamefont
  {Cheng}, \citenamefont {Gottscholl}, \citenamefont {Long}, \citenamefont
  {Mellor}, \citenamefont {Zettl}, \citenamefont {Dyakonov}, \citenamefont
  {Beton} \emph {et~al.}}]{mendelson2021identifying}%
  \BibitemOpen
  \bibfield  {author} {\bibinfo {author} {\bibfnamefont {N.}~\bibnamefont
  {Mendelson}}, \bibinfo {author} {\bibfnamefont {D.}~\bibnamefont {Chugh}},
  \bibinfo {author} {\bibfnamefont {J.~R.}\ \bibnamefont {Reimers}}, \bibinfo
  {author} {\bibfnamefont {T.~S.}\ \bibnamefont {Cheng}}, \bibinfo {author}
  {\bibfnamefont {A.}~\bibnamefont {Gottscholl}}, \bibinfo {author}
  {\bibfnamefont {H.}~\bibnamefont {Long}}, \bibinfo {author} {\bibfnamefont
  {C.~J.}\ \bibnamefont {Mellor}}, \bibinfo {author} {\bibfnamefont
  {A.}~\bibnamefont {Zettl}}, \bibinfo {author} {\bibfnamefont
  {V.}~\bibnamefont {Dyakonov}}, \bibinfo {author} {\bibfnamefont {P.~H.}\
  \bibnamefont {Beton}}, \emph {et~al.},\ }\bibfield  {title} {\bibinfo {title}
  {Identifying carbon as the source of visible single-photon emission from
  hexagonal boron nitride},\ }\href@noop {} {\bibfield  {journal} {\bibinfo
  {journal} {Nature materials}\ }\textbf {\bibinfo {volume} {20}},\ \bibinfo
  {pages} {321} (\bibinfo {year} {2021})}\BibitemShut {NoStop}%
\bibitem [{\citenamefont {Gottscholl}\ \emph {et~al.}(2020)\citenamefont
  {Gottscholl}, \citenamefont {Kianinia}, \citenamefont {Soltamov},
  \citenamefont {Orlinskii}, \citenamefont {Mamin}, \citenamefont {Bradac},
  \citenamefont {Kasper}, \citenamefont {Krambrock}, \citenamefont {Sperlich},
  \citenamefont {Toth} \emph {et~al.}}]{gottscholl2020initialization}%
  \BibitemOpen
  \bibfield  {author} {\bibinfo {author} {\bibfnamefont {A.}~\bibnamefont
  {Gottscholl}}, \bibinfo {author} {\bibfnamefont {M.}~\bibnamefont
  {Kianinia}}, \bibinfo {author} {\bibfnamefont {V.}~\bibnamefont {Soltamov}},
  \bibinfo {author} {\bibfnamefont {S.}~\bibnamefont {Orlinskii}}, \bibinfo
  {author} {\bibfnamefont {G.}~\bibnamefont {Mamin}}, \bibinfo {author}
  {\bibfnamefont {C.}~\bibnamefont {Bradac}}, \bibinfo {author} {\bibfnamefont
  {C.}~\bibnamefont {Kasper}}, \bibinfo {author} {\bibfnamefont
  {K.}~\bibnamefont {Krambrock}}, \bibinfo {author} {\bibfnamefont
  {A.}~\bibnamefont {Sperlich}}, \bibinfo {author} {\bibfnamefont
  {M.}~\bibnamefont {Toth}}, \emph {et~al.},\ }\bibfield  {title} {\bibinfo
  {title} {Initialization and read-out of intrinsic spin defects in a van der
  waals crystal at room temperature},\ }\href@noop {} {\bibfield  {journal}
  {\bibinfo  {journal} {Nature materials}\ }\textbf {\bibinfo {volume} {19}},\
  \bibinfo {pages} {540} (\bibinfo {year} {2020})}\BibitemShut {NoStop}%
\bibitem [{\citenamefont {Hoese}\ \emph {et~al.}(2020)\citenamefont {Hoese},
  \citenamefont {Reddy}, \citenamefont {Dietrich}, \citenamefont {Koch},
  \citenamefont {Fehler}, \citenamefont {Doherty},\ and\ \citenamefont
  {Kubanek}}]{hoese2020mechanical}%
  \BibitemOpen
  \bibfield  {author} {\bibinfo {author} {\bibfnamefont {M.}~\bibnamefont
  {Hoese}}, \bibinfo {author} {\bibfnamefont {P.}~\bibnamefont {Reddy}},
  \bibinfo {author} {\bibfnamefont {A.}~\bibnamefont {Dietrich}}, \bibinfo
  {author} {\bibfnamefont {M.~K.}\ \bibnamefont {Koch}}, \bibinfo {author}
  {\bibfnamefont {K.~G.}\ \bibnamefont {Fehler}}, \bibinfo {author}
  {\bibfnamefont {M.~W.}\ \bibnamefont {Doherty}},\ and\ \bibinfo {author}
  {\bibfnamefont {A.}~\bibnamefont {Kubanek}},\ }\bibfield  {title} {\bibinfo
  {title} {Mechanical decoupling of quantum emitters in hexagonal boron nitride
  from low-energy phonon modes},\ }\href@noop {} {\bibfield  {journal}
  {\bibinfo  {journal} {Science advances}\ }\textbf {\bibinfo {volume} {6}},\
  \bibinfo {pages} {eaba6038} (\bibinfo {year} {2020})}\BibitemShut {NoStop}%
\bibitem [{\citenamefont {Cassabois}\ \emph {et~al.}(2016)\citenamefont
  {Cassabois}, \citenamefont {Valvin},\ and\ \citenamefont
  {Gil}}]{cassabois2016hexagonal}%
  \BibitemOpen
  \bibfield  {author} {\bibinfo {author} {\bibfnamefont {G.}~\bibnamefont
  {Cassabois}}, \bibinfo {author} {\bibfnamefont {P.}~\bibnamefont {Valvin}},\
  and\ \bibinfo {author} {\bibfnamefont {B.}~\bibnamefont {Gil}},\ }\bibfield
  {title} {\bibinfo {title} {Hexagonal boron nitride is an indirect bandgap
  semiconductor},\ }\href@noop {} {\bibfield  {journal} {\bibinfo  {journal}
  {Nature photonics}\ }\textbf {\bibinfo {volume} {10}},\ \bibinfo {pages}
  {262} (\bibinfo {year} {2016})}\BibitemShut {NoStop}%
\bibitem [{\citenamefont {Tran}\ \emph {et~al.}(2019)\citenamefont {Tran},
  \citenamefont {Bradac}, \citenamefont {Solntsev}, \citenamefont {Toth},\ and\
  \citenamefont {Aharonovich}}]{tran2019suppression}%
  \BibitemOpen
  \bibfield  {author} {\bibinfo {author} {\bibfnamefont {T.~T.}\ \bibnamefont
  {Tran}}, \bibinfo {author} {\bibfnamefont {C.}~\bibnamefont {Bradac}},
  \bibinfo {author} {\bibfnamefont {A.~S.}\ \bibnamefont {Solntsev}}, \bibinfo
  {author} {\bibfnamefont {M.}~\bibnamefont {Toth}},\ and\ \bibinfo {author}
  {\bibfnamefont {I.}~\bibnamefont {Aharonovich}},\ }\bibfield  {title}
  {\bibinfo {title} {Suppression of spectral diffusion by anti-stokes
  excitation of quantum emitters in hexagonal boron nitride},\ }\href@noop {}
  {\bibfield  {journal} {\bibinfo  {journal} {Applied Physics Letters}\
  }\textbf {\bibinfo {volume} {115}},\ \bibinfo {pages} {071102} (\bibinfo
  {year} {2019})}\BibitemShut {NoStop}%
\bibitem [{\citenamefont {Li}\ \emph {et~al.}(2021)\citenamefont {Li},
  \citenamefont {Smart},\ and\ \citenamefont {Ping}}]{li2021c_2c_n}%
  \BibitemOpen
  \bibfield  {author} {\bibinfo {author} {\bibfnamefont {K.}~\bibnamefont
  {Li}}, \bibinfo {author} {\bibfnamefont {T.}~\bibnamefont {Smart}},\ and\
  \bibinfo {author} {\bibfnamefont {Y.}~\bibnamefont {Ping}},\ }\bibfield
  {title} {\bibinfo {title} {$\mathrm{C}_2\mathrm{C}_\mathrm{N}$ as a 2 ev
  single-photon emitter candidate in hexagonal boron nitride},\ }\href@noop {}
  {\bibfield  {journal} {\bibinfo  {journal} {arXiv preprint arXiv:2110.01787}\
  } (\bibinfo {year} {2021})}\BibitemShut {NoStop}%
\bibitem [{\citenamefont {Vogl}\ \emph {et~al.}(2019)\citenamefont {Vogl},
  \citenamefont {Lecamwasam}, \citenamefont {Buchler}, \citenamefont {Lu},\
  and\ \citenamefont {Lam}}]{vogl2019compact}%
  \BibitemOpen
  \bibfield  {author} {\bibinfo {author} {\bibfnamefont {T.}~\bibnamefont
  {Vogl}}, \bibinfo {author} {\bibfnamefont {R.}~\bibnamefont {Lecamwasam}},
  \bibinfo {author} {\bibfnamefont {B.~C.}\ \bibnamefont {Buchler}}, \bibinfo
  {author} {\bibfnamefont {Y.}~\bibnamefont {Lu}},\ and\ \bibinfo {author}
  {\bibfnamefont {P.~K.}\ \bibnamefont {Lam}},\ }\bibfield  {title} {\bibinfo
  {title} {Compact cavity-enhanced single-photon generation with hexagonal
  boron nitride},\ }\href@noop {} {\bibfield  {journal} {\bibinfo  {journal}
  {ACS Photonics}\ }\textbf {\bibinfo {volume} {6}},\ \bibinfo {pages} {1955}
  (\bibinfo {year} {2019})}\BibitemShut {NoStop}%
\bibitem [{\citenamefont {H{\"a}u{\ss}ler}\ \emph {et~al.}(2021)\citenamefont
  {H{\"a}u{\ss}ler}, \citenamefont {Bayer}, \citenamefont {Waltrich},
  \citenamefont {Mendelson}, \citenamefont {Li}, \citenamefont {Hunger},
  \citenamefont {Aharonovich},\ and\ \citenamefont
  {Kubanek}}]{haussler2021tunable}%
  \BibitemOpen
  \bibfield  {author} {\bibinfo {author} {\bibfnamefont {S.}~\bibnamefont
  {H{\"a}u{\ss}ler}}, \bibinfo {author} {\bibfnamefont {G.}~\bibnamefont
  {Bayer}}, \bibinfo {author} {\bibfnamefont {R.}~\bibnamefont {Waltrich}},
  \bibinfo {author} {\bibfnamefont {N.}~\bibnamefont {Mendelson}}, \bibinfo
  {author} {\bibfnamefont {C.}~\bibnamefont {Li}}, \bibinfo {author}
  {\bibfnamefont {D.}~\bibnamefont {Hunger}}, \bibinfo {author} {\bibfnamefont
  {I.}~\bibnamefont {Aharonovich}},\ and\ \bibinfo {author} {\bibfnamefont
  {A.}~\bibnamefont {Kubanek}},\ }\bibfield  {title} {\bibinfo {title} {Tunable
  fiber-cavity enhanced photon emission from defect centers in hbn},\
  }\href@noop {} {\bibfield  {journal} {\bibinfo  {journal} {Advanced Optical
  Materials}\ }\textbf {\bibinfo {volume} {9}},\ \bibinfo {pages} {2002218}
  (\bibinfo {year} {2021})}\BibitemShut {NoStop}%
\bibitem [{\citenamefont {Denning}\ \emph {et~al.}(2020)\citenamefont
  {Denning}, \citenamefont {Iles-Smith}, \citenamefont {Gregersen},\ and\
  \citenamefont {Mork}}]{Denning20phonons}%
  \BibitemOpen
  \bibfield  {author} {\bibinfo {author} {\bibfnamefont {E.~V.}\ \bibnamefont
  {Denning}}, \bibinfo {author} {\bibfnamefont {J.}~\bibnamefont {Iles-Smith}},
  \bibinfo {author} {\bibfnamefont {N.}~\bibnamefont {Gregersen}},\ and\
  \bibinfo {author} {\bibfnamefont {J.}~\bibnamefont {Mork}},\ }\bibfield
  {title} {\bibinfo {title} {Phonon effects in quantum dot single-photon
  sources},\ }\href {https://doi.org/10.1364/OME.380601} {\bibfield  {journal}
  {\bibinfo  {journal} {Opt. Mater. Express}\ }\textbf {\bibinfo {volume}
  {10}},\ \bibinfo {pages} {222} (\bibinfo {year} {2020})}\BibitemShut
  {NoStop}%
\bibitem [{\citenamefont {Nazir}\ and\ \citenamefont
  {McCutcheon}(2016)}]{Nazir16review}%
  \BibitemOpen
  \bibfield  {author} {\bibinfo {author} {\bibfnamefont {A.}~\bibnamefont
  {Nazir}}\ and\ \bibinfo {author} {\bibfnamefont {D.~P.~S.}\ \bibnamefont
  {McCutcheon}},\ }\bibfield  {title} {\bibinfo {title} {Modelling
  exciton{\textendash}phonon interactions in optically driven quantum dots},\
  }\href {https://doi.org/10.1088/0953-8984/28/10/103002} {\bibfield  {journal}
  {\bibinfo  {journal} {Journal of Physics: Condensed Matter}\ }\textbf
  {\bibinfo {volume} {28}},\ \bibinfo {pages} {103002} (\bibinfo {year}
  {2016})}\BibitemShut {NoStop}%
\bibitem [{\citenamefont {Vuong}\ \emph {et~al.}(2016)\citenamefont {Vuong},
  \citenamefont {Cassabois}, \citenamefont {Valvin}, \citenamefont {Ouerghi},
  \citenamefont {Chassagneux}, \citenamefont {Voisin},\ and\ \citenamefont
  {Gil}}]{vuong2016phonon}%
  \BibitemOpen
  \bibfield  {author} {\bibinfo {author} {\bibfnamefont {T.}~\bibnamefont
  {Vuong}}, \bibinfo {author} {\bibfnamefont {G.}~\bibnamefont {Cassabois}},
  \bibinfo {author} {\bibfnamefont {P.}~\bibnamefont {Valvin}}, \bibinfo
  {author} {\bibfnamefont {A.}~\bibnamefont {Ouerghi}}, \bibinfo {author}
  {\bibfnamefont {Y.}~\bibnamefont {Chassagneux}}, \bibinfo {author}
  {\bibfnamefont {C.}~\bibnamefont {Voisin}},\ and\ \bibinfo {author}
  {\bibfnamefont {B.}~\bibnamefont {Gil}},\ }\bibfield  {title} {\bibinfo
  {title} {Phonon-photon mapping in a color center in hexagonal boron
  nitride},\ }\href@noop {} {\bibfield  {journal} {\bibinfo  {journal}
  {Physical review letters}\ }\textbf {\bibinfo {volume} {117}},\ \bibinfo
  {pages} {097402} (\bibinfo {year} {2016})}\BibitemShut {NoStop}%
\bibitem [{\citenamefont {Feldman}\ \emph {et~al.}(2019)\citenamefont
  {Feldman}, \citenamefont {Puretzky}, \citenamefont {Lindsay}, \citenamefont
  {Tucker}, \citenamefont {Briggs}, \citenamefont {Evans}, \citenamefont
  {Haglund},\ and\ \citenamefont {Lawrie}}]{Feldman19phonon}%
  \BibitemOpen
  \bibfield  {author} {\bibinfo {author} {\bibfnamefont {M.~A.}\ \bibnamefont
  {Feldman}}, \bibinfo {author} {\bibfnamefont {A.}~\bibnamefont {Puretzky}},
  \bibinfo {author} {\bibfnamefont {L.}~\bibnamefont {Lindsay}}, \bibinfo
  {author} {\bibfnamefont {E.}~\bibnamefont {Tucker}}, \bibinfo {author}
  {\bibfnamefont {D.~P.}\ \bibnamefont {Briggs}}, \bibinfo {author}
  {\bibfnamefont {P.~G.}\ \bibnamefont {Evans}}, \bibinfo {author}
  {\bibfnamefont {R.~F.}\ \bibnamefont {Haglund}},\ and\ \bibinfo {author}
  {\bibfnamefont {B.~J.}\ \bibnamefont {Lawrie}},\ }\bibfield  {title}
  {\bibinfo {title} {Phonon-induced multicolor correlations in hbn
  single-photon emitters},\ }\href {https://doi.org/10.1103/PhysRevB.99.020101}
  {\bibfield  {journal} {\bibinfo  {journal} {Phys. Rev. B}\ }\textbf {\bibinfo
  {volume} {99}},\ \bibinfo {pages} {020101} (\bibinfo {year}
  {2019})}\BibitemShut {NoStop}%
\bibitem [{\citenamefont {Khatri}\ \emph {et~al.}(2019)\citenamefont {Khatri},
  \citenamefont {Luxmoore},\ and\ \citenamefont {Ramsay}}]{Khatri19sideband}%
  \BibitemOpen
  \bibfield  {author} {\bibinfo {author} {\bibfnamefont {P.}~\bibnamefont
  {Khatri}}, \bibinfo {author} {\bibfnamefont {I.~J.}\ \bibnamefont
  {Luxmoore}},\ and\ \bibinfo {author} {\bibfnamefont {A.~J.}\ \bibnamefont
  {Ramsay}},\ }\bibfield  {title} {\bibinfo {title} {Phonon sidebands of color
  centers in hexagonal boron nitride},\ }\href
  {https://doi.org/10.1103/PhysRevB.100.125305} {\bibfield  {journal} {\bibinfo
   {journal} {Phys. Rev. B}\ }\textbf {\bibinfo {volume} {100}},\ \bibinfo
  {pages} {125305} (\bibinfo {year} {2019})}\BibitemShut {NoStop}%
\bibitem [{\citenamefont {Grosso}\ \emph {et~al.}(2020)\citenamefont {Grosso},
  \citenamefont {Moon}, \citenamefont {Ciccarino}, \citenamefont {Flick},
  \citenamefont {Mendelson}, \citenamefont {Mennel}, \citenamefont {Toth},
  \citenamefont {Aharonovich}, \citenamefont {Narang},\ and\ \citenamefont
  {Englund}}]{Grosso20lowtemp}%
  \BibitemOpen
  \bibfield  {author} {\bibinfo {author} {\bibfnamefont {G.}~\bibnamefont
  {Grosso}}, \bibinfo {author} {\bibfnamefont {H.}~\bibnamefont {Moon}},
  \bibinfo {author} {\bibfnamefont {C.~J.}\ \bibnamefont {Ciccarino}}, \bibinfo
  {author} {\bibfnamefont {J.}~\bibnamefont {Flick}}, \bibinfo {author}
  {\bibfnamefont {N.}~\bibnamefont {Mendelson}}, \bibinfo {author}
  {\bibfnamefont {L.}~\bibnamefont {Mennel}}, \bibinfo {author} {\bibfnamefont
  {M.}~\bibnamefont {Toth}}, \bibinfo {author} {\bibfnamefont {I.}~\bibnamefont
  {Aharonovich}}, \bibinfo {author} {\bibfnamefont {P.}~\bibnamefont
  {Narang}},\ and\ \bibinfo {author} {\bibfnamefont {D.~R.}\ \bibnamefont
  {Englund}},\ }\bibfield  {title} {\bibinfo {title} {Low-temperature
  electron–phonon interaction of quantum emitters in hexagonal boron
  nitride},\ }\href {https://doi.org/10.1021/acsphotonics.9b01789} {\bibfield
  {journal} {\bibinfo  {journal} {ACS Photonics}\ }\textbf {\bibinfo {volume}
  {7}},\ \bibinfo {pages} {1410} (\bibinfo {year} {2020})}\BibitemShut
  {NoStop}%
\bibitem [{\citenamefont {Sajid}\ and\ \citenamefont
  {Thygesen}(2020)}]{sajid2020vncb}%
  \BibitemOpen
  \bibfield  {author} {\bibinfo {author} {\bibfnamefont {A.}~\bibnamefont
  {Sajid}}\ and\ \bibinfo {author} {\bibfnamefont {K.~S.}\ \bibnamefont
  {Thygesen}},\ }\bibfield  {title} {\bibinfo {title} {Vncb defect as source of
  single photon emission from hexagonal boron nitride},\ }\href@noop {}
  {\bibfield  {journal} {\bibinfo  {journal} {2D Materials}\ }\textbf {\bibinfo
  {volume} {7}},\ \bibinfo {pages} {031007} (\bibinfo {year}
  {2020})}\BibitemShut {NoStop}%
\bibitem [{\citenamefont {Jara}\ \emph {et~al.}(2021)\citenamefont {Jara},
  \citenamefont {Rauch}, \citenamefont {Botti}, \citenamefont {Marques},
  \citenamefont {Norambuena}, \citenamefont {Coto}, \citenamefont
  {Castellanos-{\'A}guila}, \citenamefont {Maze},\ and\ \citenamefont
  {Munoz}}]{jara2021first}%
  \BibitemOpen
  \bibfield  {author} {\bibinfo {author} {\bibfnamefont {C.}~\bibnamefont
  {Jara}}, \bibinfo {author} {\bibfnamefont {T.}~\bibnamefont {Rauch}},
  \bibinfo {author} {\bibfnamefont {S.}~\bibnamefont {Botti}}, \bibinfo
  {author} {\bibfnamefont {M.~A.}\ \bibnamefont {Marques}}, \bibinfo {author}
  {\bibfnamefont {A.}~\bibnamefont {Norambuena}}, \bibinfo {author}
  {\bibfnamefont {R.}~\bibnamefont {Coto}}, \bibinfo {author} {\bibfnamefont
  {J.}~\bibnamefont {Castellanos-{\'A}guila}}, \bibinfo {author} {\bibfnamefont
  {J.~R.}\ \bibnamefont {Maze}},\ and\ \bibinfo {author} {\bibfnamefont
  {F.}~\bibnamefont {Munoz}},\ }\bibfield  {title} {\bibinfo {title}
  {First-principles identification of single photon emitters based on carbon
  clusters in hexagonal boron nitride},\ }\href@noop {} {\bibfield  {journal}
  {\bibinfo  {journal} {The Journal of Physical Chemistry A}\ }\textbf
  {\bibinfo {volume} {125}},\ \bibinfo {pages} {1325} (\bibinfo {year}
  {2021})}\BibitemShut {NoStop}%
\bibitem [{\citenamefont {Li}\ \emph {et~al.}(2022)\citenamefont {Li},
  \citenamefont {Smart},\ and\ \citenamefont {Ping}}]{li2022carbon}%
  \BibitemOpen
  \bibfield  {author} {\bibinfo {author} {\bibfnamefont {K.}~\bibnamefont
  {Li}}, \bibinfo {author} {\bibfnamefont {T.~J.}\ \bibnamefont {Smart}},\ and\
  \bibinfo {author} {\bibfnamefont {Y.}~\bibnamefont {Ping}},\ }\bibfield
  {title} {\bibinfo {title} {Carbon trimer as a 2 ev single-photon emitter
  candidate in hexagonal boron nitride: A first-principles study},\ }\href@noop
  {} {\bibfield  {journal} {\bibinfo  {journal} {Physical Review Materials}\
  }\textbf {\bibinfo {volume} {6}},\ \bibinfo {pages} {L042201} (\bibinfo
  {year} {2022})}\BibitemShut {NoStop}%
\bibitem [{\citenamefont {Strathearn}\ \emph {et~al.}(2018)\citenamefont
  {Strathearn}, \citenamefont {Kirton}, \citenamefont {Kilda}, \citenamefont
  {Keeling},\ and\ \citenamefont {Lovett}}]{strathearn2018efficient}%
  \BibitemOpen
  \bibfield  {author} {\bibinfo {author} {\bibfnamefont {A.}~\bibnamefont
  {Strathearn}}, \bibinfo {author} {\bibfnamefont {P.}~\bibnamefont {Kirton}},
  \bibinfo {author} {\bibfnamefont {D.}~\bibnamefont {Kilda}}, \bibinfo
  {author} {\bibfnamefont {J.}~\bibnamefont {Keeling}},\ and\ \bibinfo {author}
  {\bibfnamefont {B.~W.}\ \bibnamefont {Lovett}},\ }\bibfield  {title}
  {\bibinfo {title} {Efficient non-markovian quantum dynamics using
  time-evolving matrix product operators},\ }\href@noop {} {\bibfield
  {journal} {\bibinfo  {journal} {Nature communications}\ }\textbf {\bibinfo
  {volume} {9}},\ \bibinfo {pages} {1} (\bibinfo {year} {2018})}\BibitemShut
  {NoStop}%
\bibitem [{\citenamefont {J\o{}rgensen}\ and\ \citenamefont
  {Pollock}(2019)}]{Jorgensen19causal}%
  \BibitemOpen
  \bibfield  {author} {\bibinfo {author} {\bibfnamefont {M.~R.}\ \bibnamefont
  {J\o{}rgensen}}\ and\ \bibinfo {author} {\bibfnamefont {F.~A.}\ \bibnamefont
  {Pollock}},\ }\bibfield  {title} {\bibinfo {title} {Exploiting the causal
  tensor network structure of quantum processes to efficiently simulate
  non-markovian path integrals},\ }\href
  {https://doi.org/10.1103/PhysRevLett.123.240602} {\bibfield  {journal}
  {\bibinfo  {journal} {Phys. Rev. Lett.}\ }\textbf {\bibinfo {volume} {123}},\
  \bibinfo {pages} {240602} (\bibinfo {year} {2019})}\BibitemShut {NoStop}%
\bibitem [{\citenamefont {Gribben}\ \emph {et~al.}(2020)\citenamefont
  {Gribben}, \citenamefont {Strathearn}, \citenamefont {Iles-Smith},
  \citenamefont {Kilda}, \citenamefont {Nazir}, \citenamefont {Lovett},\ and\
  \citenamefont {Kirton}}]{Gribben20structure}%
  \BibitemOpen
  \bibfield  {author} {\bibinfo {author} {\bibfnamefont {D.}~\bibnamefont
  {Gribben}}, \bibinfo {author} {\bibfnamefont {A.}~\bibnamefont {Strathearn}},
  \bibinfo {author} {\bibfnamefont {J.}~\bibnamefont {Iles-Smith}}, \bibinfo
  {author} {\bibfnamefont {D.}~\bibnamefont {Kilda}}, \bibinfo {author}
  {\bibfnamefont {A.}~\bibnamefont {Nazir}}, \bibinfo {author} {\bibfnamefont
  {B.~W.}\ \bibnamefont {Lovett}},\ and\ \bibinfo {author} {\bibfnamefont
  {P.}~\bibnamefont {Kirton}},\ }\bibfield  {title} {\bibinfo {title} {Exact
  quantum dynamics in structured environments},\ }\href
  {https://doi.org/10.1103/PhysRevResearch.2.013265} {\bibfield  {journal}
  {\bibinfo  {journal} {Phys. Rev. Research}\ }\textbf {\bibinfo {volume}
  {2}},\ \bibinfo {pages} {013265} (\bibinfo {year} {2020})}\BibitemShut
  {NoStop}%
\bibitem [{\citenamefont {Gribben}\ \emph {et~al.}(2022)\citenamefont
  {Gribben}, \citenamefont {Rouse}, \citenamefont {Iles-Smith}, \citenamefont
  {Strathearn}, \citenamefont {Maguire}, \citenamefont {Kirton}, \citenamefont
  {Nazir}, \citenamefont {Gauger},\ and\ \citenamefont
  {Lovett}}]{Gribben22exact}%
  \BibitemOpen
  \bibfield  {author} {\bibinfo {author} {\bibfnamefont {D.}~\bibnamefont
  {Gribben}}, \bibinfo {author} {\bibfnamefont {D.~M.}\ \bibnamefont {Rouse}},
  \bibinfo {author} {\bibfnamefont {J.}~\bibnamefont {Iles-Smith}}, \bibinfo
  {author} {\bibfnamefont {A.}~\bibnamefont {Strathearn}}, \bibinfo {author}
  {\bibfnamefont {H.}~\bibnamefont {Maguire}}, \bibinfo {author} {\bibfnamefont
  {P.}~\bibnamefont {Kirton}}, \bibinfo {author} {\bibfnamefont
  {A.}~\bibnamefont {Nazir}}, \bibinfo {author} {\bibfnamefont {E.~M.}\
  \bibnamefont {Gauger}},\ and\ \bibinfo {author} {\bibfnamefont {B.~W.}\
  \bibnamefont {Lovett}},\ }\bibfield  {title} {\bibinfo {title} {Exact
  dynamics of nonadditive environments in non-markovian open quantum systems},\
  }\href {https://doi.org/10.1103/PRXQuantum.3.010321} {\bibfield  {journal}
  {\bibinfo  {journal} {PRX Quantum}\ }\textbf {\bibinfo {volume} {3}},\
  \bibinfo {pages} {010321} (\bibinfo {year} {2022})}\BibitemShut {NoStop}%
\bibitem [{\citenamefont {Svendsen}\ \emph {et~al.}(2021)\citenamefont
  {Svendsen}, \citenamefont {Kurman} \emph {et~al.}}]{svendsen2021combining}%
  \BibitemOpen
  \bibfield  {author} {\bibinfo {author} {\bibfnamefont {M.~K.}\ \bibnamefont
  {Svendsen}}, \bibinfo {author} {\bibfnamefont {Y.}~\bibnamefont {Kurman}},
  \emph {et~al.},\ }\bibfield  {title} {\bibinfo {title} {Combining density
  functional theory with macroscopic qed for quantum light-matter interactions
  in 2$\mathrm{D}$ materials},\ }\href@noop {} {\bibfield  {journal} {\bibinfo
  {journal} {Nature communications}\ }\textbf {\bibinfo {volume} {12}},\
  \bibinfo {pages} {1} (\bibinfo {year} {2021})}\BibitemShut {NoStop}%
\bibitem [{\citenamefont {Latini}\ \emph {et~al.}(2021)\citenamefont {Latini},
  \citenamefont {De~Giovannini}, \citenamefont {Sie}, \citenamefont {Gedik},
  \citenamefont {H\"ubener},\ and\ \citenamefont
  {Rubio}}]{Latini21phonoritons}%
  \BibitemOpen
  \bibfield  {author} {\bibinfo {author} {\bibfnamefont {S.}~\bibnamefont
  {Latini}}, \bibinfo {author} {\bibfnamefont {U.}~\bibnamefont
  {De~Giovannini}}, \bibinfo {author} {\bibfnamefont {E.~J.}\ \bibnamefont
  {Sie}}, \bibinfo {author} {\bibfnamefont {N.}~\bibnamefont {Gedik}}, \bibinfo
  {author} {\bibfnamefont {H.}~\bibnamefont {H\"ubener}},\ and\ \bibinfo
  {author} {\bibfnamefont {A.}~\bibnamefont {Rubio}},\ }\bibfield  {title}
  {\bibinfo {title} {Phonoritons as hybridized exciton-photon-phonon
  excitations in a monolayer $\mathrm{h-BN}$ optical cavity},\ }\href
  {https://doi.org/10.1103/PhysRevLett.126.227401} {\bibfield  {journal}
  {\bibinfo  {journal} {Phys. Rev. Lett.}\ }\textbf {\bibinfo {volume} {126}},\
  \bibinfo {pages} {227401} (\bibinfo {year} {2021})}\BibitemShut {NoStop}%
\bibitem [{\citenamefont {Sajid}\ \emph {et~al.}(2018)\citenamefont {Sajid},
  \citenamefont {Reimers},\ and\ \citenamefont {Ford}}]{Sajid18defect}%
  \BibitemOpen
  \bibfield  {author} {\bibinfo {author} {\bibfnamefont {A.}~\bibnamefont
  {Sajid}}, \bibinfo {author} {\bibfnamefont {J.~R.}\ \bibnamefont {Reimers}},\
  and\ \bibinfo {author} {\bibfnamefont {M.~J.}\ \bibnamefont {Ford}},\
  }\bibfield  {title} {\bibinfo {title} {Defect states in hexagonal boron
  nitride: Assignments of observed properties and prediction of properties
  relevant to quantum computation},\ }\href
  {https://doi.org/10.1103/PhysRevB.97.064101} {\bibfield  {journal} {\bibinfo
  {journal} {Phys. Rev. B}\ }\textbf {\bibinfo {volume} {97}},\ \bibinfo
  {pages} {064101} (\bibinfo {year} {2018})}\BibitemShut {NoStop}%
\bibitem [{Note1()}]{Note1}%
  \BibitemOpen
  \bibinfo {note} {This analysis is repeated in the supplementary information
  for the C$_2$C$_N$ defect.}\BibitemShut {Stop}%
\bibitem [{\citenamefont {Carmichael}(1999)}]{carmichael1999statistical}%
  \BibitemOpen
  \bibfield  {author} {\bibinfo {author} {\bibfnamefont {H.~J.}\ \bibnamefont
  {Carmichael}},\ }\href@noop {} {\emph {\bibinfo {title} {Statistical methods
  in quantum optics 1: master equations and Fokker-Planck equations}}},\
  Vol.~\bibinfo {volume} {1}\ (\bibinfo  {publisher} {Springer Science \&
  Business Media},\ \bibinfo {year} {1999})\BibitemShut {NoStop}%
\bibitem [{\citenamefont {Maguire}\ \emph {et~al.}(2019)\citenamefont
  {Maguire}, \citenamefont {Iles-Smith},\ and\ \citenamefont
  {Nazir}}]{maguire2019environmental}%
  \BibitemOpen
  \bibfield  {author} {\bibinfo {author} {\bibfnamefont {H.}~\bibnamefont
  {Maguire}}, \bibinfo {author} {\bibfnamefont {J.}~\bibnamefont
  {Iles-Smith}},\ and\ \bibinfo {author} {\bibfnamefont {A.}~\bibnamefont
  {Nazir}},\ }\bibfield  {title} {\bibinfo {title} {Environmental nonadditivity
  and franck-condon physics in nonequilibrium quantum systems},\ }\href@noop {}
  {\bibfield  {journal} {\bibinfo  {journal} {Physical review letters}\
  }\textbf {\bibinfo {volume} {123}},\ \bibinfo {pages} {093601} (\bibinfo
  {year} {2019})}\BibitemShut {NoStop}%
\bibitem [{\citenamefont {Mahan}(2013)}]{mahan2013many}%
  \BibitemOpen
  \bibfield  {author} {\bibinfo {author} {\bibfnamefont {G.~D.}\ \bibnamefont
  {Mahan}},\ }\href@noop {} {\emph {\bibinfo {title} {Many-particle physics}}}\
  (\bibinfo  {publisher} {Springer Science \& Business Media},\ \bibinfo {year}
  {2013})\BibitemShut {NoStop}%
\bibitem [{\citenamefont {Borrelli}\ \emph {et~al.}(2013)\citenamefont
  {Borrelli}, \citenamefont {Capobianco},\ and\ \citenamefont
  {Peluso}}]{borrelli2013franck}%
  \BibitemOpen
  \bibfield  {author} {\bibinfo {author} {\bibfnamefont {R.}~\bibnamefont
  {Borrelli}}, \bibinfo {author} {\bibfnamefont {A.}~\bibnamefont
  {Capobianco}},\ and\ \bibinfo {author} {\bibfnamefont {A.}~\bibnamefont
  {Peluso}},\ }\bibfield  {title} {\bibinfo {title} {Franck--condon
  factors—computational approaches and recent developments},\ }\href@noop {}
  {\bibfield  {journal} {\bibinfo  {journal} {Canadian Journal of Chemistry}\
  }\textbf {\bibinfo {volume} {91}},\ \bibinfo {pages} {495} (\bibinfo {year}
  {2013})}\BibitemShut {NoStop}%
\bibitem [{\citenamefont {Alkauskas}\ \emph {et~al.}(2014)\citenamefont
  {Alkauskas}, \citenamefont {Buckley}, \citenamefont {Awschalom},\ and\
  \citenamefont {Van~de Walle}}]{alkauskas2014first}%
  \BibitemOpen
  \bibfield  {author} {\bibinfo {author} {\bibfnamefont {A.}~\bibnamefont
  {Alkauskas}}, \bibinfo {author} {\bibfnamefont {B.~B.}\ \bibnamefont
  {Buckley}}, \bibinfo {author} {\bibfnamefont {D.~D.}\ \bibnamefont
  {Awschalom}},\ and\ \bibinfo {author} {\bibfnamefont {C.~G.}\ \bibnamefont
  {Van~de Walle}},\ }\bibfield  {title} {\bibinfo {title} {First-principles
  theory of the luminescence lineshape for the triplet transition in diamond nv
  centres},\ }\href@noop {} {\bibfield  {journal} {\bibinfo  {journal} {New
  Journal of Physics}\ }\textbf {\bibinfo {volume} {16}},\ \bibinfo {pages}
  {073026} (\bibinfo {year} {2014})}\BibitemShut {NoStop}%
\bibitem [{\citenamefont {Kresse}\ and\ \citenamefont
  {Hafner}(1993)}]{kresse1993ab}%
  \BibitemOpen
  \bibfield  {author} {\bibinfo {author} {\bibfnamefont {G.}~\bibnamefont
  {Kresse}}\ and\ \bibinfo {author} {\bibfnamefont {J.}~\bibnamefont
  {Hafner}},\ }\bibfield  {title} {\bibinfo {title} {Ab initio molecular
  dynamics for liquid metals},\ }\href@noop {} {\bibfield  {journal} {\bibinfo
  {journal} {Physical review B}\ }\textbf {\bibinfo {volume} {47}},\ \bibinfo
  {pages} {558} (\bibinfo {year} {1993})}\BibitemShut {NoStop}%
\bibitem [{\citenamefont {Heyd}\ and\ \citenamefont
  {Scuseria}(2004)}]{heyd2004efficient}%
  \BibitemOpen
  \bibfield  {author} {\bibinfo {author} {\bibfnamefont {J.}~\bibnamefont
  {Heyd}}\ and\ \bibinfo {author} {\bibfnamefont {G.~E.}\ \bibnamefont
  {Scuseria}},\ }\bibfield  {title} {\bibinfo {title} {Efficient hybrid density
  functional calculations in solids: Assessment of the
  heyd--scuseria--ernzerhof screened coulomb hybrid functional},\ }\href@noop
  {} {\bibfield  {journal} {\bibinfo  {journal} {The Journal of chemical
  physics}\ }\textbf {\bibinfo {volume} {121}},\ \bibinfo {pages} {1187}
  (\bibinfo {year} {2004})}\BibitemShut {NoStop}%
\bibitem [{\citenamefont {Breuer}\ \emph {et~al.}(2002)\citenamefont {Breuer},
  \citenamefont {Petruccione} \emph {et~al.}}]{breuer2002theory}%
  \BibitemOpen
  \bibfield  {author} {\bibinfo {author} {\bibfnamefont {H.-P.}\ \bibnamefont
  {Breuer}}, \bibinfo {author} {\bibfnamefont {F.}~\bibnamefont {Petruccione}},
  \emph {et~al.},\ }\href@noop {} {\emph {\bibinfo {title} {The theory of open
  quantum systems}}}\ (\bibinfo  {publisher} {Oxford University Press on
  Demand},\ \bibinfo {year} {2002})\BibitemShut {NoStop}%
\bibitem [{\citenamefont {Popovic}\ \emph {et~al.}(2021)\citenamefont
  {Popovic}, \citenamefont {Mitchison}, \citenamefont {Strathearn},
  \citenamefont {Lovett}, \citenamefont {Goold},\ and\ \citenamefont
  {Eastham}}]{Popovic21heat}%
  \BibitemOpen
  \bibfield  {author} {\bibinfo {author} {\bibfnamefont {M.}~\bibnamefont
  {Popovic}}, \bibinfo {author} {\bibfnamefont {M.~T.}\ \bibnamefont
  {Mitchison}}, \bibinfo {author} {\bibfnamefont {A.}~\bibnamefont
  {Strathearn}}, \bibinfo {author} {\bibfnamefont {B.~W.}\ \bibnamefont
  {Lovett}}, \bibinfo {author} {\bibfnamefont {J.}~\bibnamefont {Goold}},\ and\
  \bibinfo {author} {\bibfnamefont {P.~R.}\ \bibnamefont {Eastham}},\
  }\bibfield  {title} {\bibinfo {title} {Quantum heat statistics with
  time-evolving matrix product operators},\ }\href
  {https://doi.org/10.1103/PRXQuantum.2.020338} {\bibfield  {journal} {\bibinfo
   {journal} {PRX Quantum}\ }\textbf {\bibinfo {volume} {2}},\ \bibinfo {pages}
  {020338} (\bibinfo {year} {2021})}\BibitemShut {NoStop}%
\bibitem [{\citenamefont {Fux}\ \emph {et~al.}(2021)\citenamefont {Fux},
  \citenamefont {Butler}, \citenamefont {Eastham}, \citenamefont {Lovett},\
  and\ \citenamefont {Keeling}}]{Fux21efficient}%
  \BibitemOpen
  \bibfield  {author} {\bibinfo {author} {\bibfnamefont {G.~E.}\ \bibnamefont
  {Fux}}, \bibinfo {author} {\bibfnamefont {E.~P.}\ \bibnamefont {Butler}},
  \bibinfo {author} {\bibfnamefont {P.~R.}\ \bibnamefont {Eastham}}, \bibinfo
  {author} {\bibfnamefont {B.~W.}\ \bibnamefont {Lovett}},\ and\ \bibinfo
  {author} {\bibfnamefont {J.}~\bibnamefont {Keeling}},\ }\bibfield  {title}
  {\bibinfo {title} {Efficient exploration of hamiltonian parameter space for
  optimal control of non-markovian open quantum systems},\ }\href
  {https://doi.org/10.1103/PhysRevLett.126.200401} {\bibfield  {journal}
  {\bibinfo  {journal} {Phys. Rev. Lett.}\ }\textbf {\bibinfo {volume} {126}},\
  \bibinfo {pages} {200401} (\bibinfo {year} {2021})}\BibitemShut {NoStop}%
\bibitem [{\citenamefont {Makri}\ and\ \citenamefont
  {Makarov}(1995{\natexlab{a}})}]{makri1995tensorI}%
  \BibitemOpen
  \bibfield  {author} {\bibinfo {author} {\bibfnamefont {N.}~\bibnamefont
  {Makri}}\ and\ \bibinfo {author} {\bibfnamefont {D.~E.}\ \bibnamefont
  {Makarov}},\ }\bibfield  {title} {\bibinfo {title} {Tensor propagator for
  iterative quantum time evolution of reduced density matrices. i. theory},\
  }\href@noop {} {\bibfield  {journal} {\bibinfo  {journal} {The Journal of
  chemical physics}\ }\textbf {\bibinfo {volume} {102}},\ \bibinfo {pages}
  {4600} (\bibinfo {year} {1995}{\natexlab{a}})}\BibitemShut {NoStop}%
\bibitem [{\citenamefont {Makri}\ and\ \citenamefont
  {Makarov}(1995{\natexlab{b}})}]{makri1995tensorII}%
  \BibitemOpen
  \bibfield  {author} {\bibinfo {author} {\bibfnamefont {N.}~\bibnamefont
  {Makri}}\ and\ \bibinfo {author} {\bibfnamefont {D.~E.}\ \bibnamefont
  {Makarov}},\ }\bibfield  {title} {\bibinfo {title} {Tensor propagator for
  iterative quantum time evolution of reduced density matrices. ii. numerical
  methodology},\ }\href@noop {} {\bibfield  {journal} {\bibinfo  {journal} {The
  Journal of chemical physics}\ }\textbf {\bibinfo {volume} {102}},\ \bibinfo
  {pages} {4611} (\bibinfo {year} {1995}{\natexlab{b}})}\BibitemShut {NoStop}%
\bibitem [{\citenamefont {Cygorek}\ \emph {et~al.}(2022)\citenamefont
  {Cygorek}, \citenamefont {Cosacchi}, \citenamefont {Vagov}, \citenamefont
  {Axt}, \citenamefont {Lovett}, \citenamefont {Keeling},\ and\ \citenamefont
  {Gauger}}]{Cygorek22ace}%
  \BibitemOpen
  \bibfield  {author} {\bibinfo {author} {\bibfnamefont {M.}~\bibnamefont
  {Cygorek}}, \bibinfo {author} {\bibfnamefont {M.}~\bibnamefont {Cosacchi}},
  \bibinfo {author} {\bibfnamefont {A.}~\bibnamefont {Vagov}}, \bibinfo
  {author} {\bibfnamefont {V.}~\bibnamefont {Axt}}, \bibinfo {author}
  {\bibfnamefont {B.}~\bibnamefont {Lovett}}, \bibinfo {author} {\bibfnamefont
  {J.}~\bibnamefont {Keeling}},\ and\ \bibinfo {author} {\bibfnamefont
  {E.}~\bibnamefont {Gauger}},\ }\bibfield  {title} {\bibinfo {title}
  {Simulation of open quantum systems by automated compression of arbitrary
  environments},\ }\href@noop {} {\bibfield  {journal} {\bibinfo  {journal}
  {Nature Physics}\ } (\bibinfo {year} {2022})}\BibitemShut {NoStop}%
\bibitem [{\citenamefont {Trotter}(1959)}]{trotter1959product}%
  \BibitemOpen
  \bibfield  {author} {\bibinfo {author} {\bibfnamefont {H.~F.}\ \bibnamefont
  {Trotter}},\ }\bibfield  {title} {\bibinfo {title} {On the product of
  semi-groups of operators},\ }\href@noop {} {\bibfield  {journal} {\bibinfo
  {journal} {Proceedings of the American Mathematical Society}\ }\textbf
  {\bibinfo {volume} {10}},\ \bibinfo {pages} {545} (\bibinfo {year}
  {1959})}\BibitemShut {NoStop}%
\bibitem [{\citenamefont {Strathearn}\ \emph {et~al.}(2017)\citenamefont
  {Strathearn}, \citenamefont {Lovett},\ and\ \citenamefont
  {Kirton}}]{Strathearn17quapi}%
  \BibitemOpen
  \bibfield  {author} {\bibinfo {author} {\bibfnamefont {A.}~\bibnamefont
  {Strathearn}}, \bibinfo {author} {\bibfnamefont {B.~W.}\ \bibnamefont
  {Lovett}},\ and\ \bibinfo {author} {\bibfnamefont {P.}~\bibnamefont
  {Kirton}},\ }\bibfield  {title} {\bibinfo {title} {Efficient real-time path
  integrals for non-markovian spin-boson models},\ }\href
  {https://doi.org/10.1088/1367-2630/aa8744} {\bibfield  {journal} {\bibinfo
  {journal} {New Journal of Physics}\ }\textbf {\bibinfo {volume} {19}},\
  \bibinfo {pages} {093009} (\bibinfo {year} {2017})}\BibitemShut {NoStop}%
\bibitem [{\citenamefont {Orús}(2014)}]{orus14intro}%
  \BibitemOpen
  \bibfield  {author} {\bibinfo {author} {\bibfnamefont {R.}~\bibnamefont
  {Orús}},\ }\bibfield  {title} {\bibinfo {title} {A practical introduction to
  tensor networks: Matrix product states and projected entangled pair states},\
  }\href {https://doi.org/https://doi.org/10.1016/j.aop.2014.06.013} {\bibfield
   {journal} {\bibinfo  {journal} {Annals of Physics}\ }\textbf {\bibinfo
  {volume} {349}},\ \bibinfo {pages} {117} (\bibinfo {year}
  {2014})}\BibitemShut {NoStop}%
\bibitem [{\citenamefont {Ruggenthaler}\ \emph {et~al.}(2018)\citenamefont
  {Ruggenthaler}, \citenamefont {Tancogne-Dejean}, \citenamefont {Flick},
  \citenamefont {Appel},\ and\ \citenamefont
  {Rubio}}]{ruggenthaler2018quantum}%
  \BibitemOpen
  \bibfield  {author} {\bibinfo {author} {\bibfnamefont {M.}~\bibnamefont
  {Ruggenthaler}}, \bibinfo {author} {\bibfnamefont {N.}~\bibnamefont
  {Tancogne-Dejean}}, \bibinfo {author} {\bibfnamefont {J.}~\bibnamefont
  {Flick}}, \bibinfo {author} {\bibfnamefont {H.}~\bibnamefont {Appel}},\ and\
  \bibinfo {author} {\bibfnamefont {A.}~\bibnamefont {Rubio}},\ }\bibfield
  {title} {\bibinfo {title} {From a quantum-electrodynamical light--matter
  description to novel spectroscopies},\ }\href@noop {} {\bibfield  {journal}
  {\bibinfo  {journal} {Nature Reviews Chemistry}\ }\textbf {\bibinfo {volume}
  {2}},\ \bibinfo {pages} {1} (\bibinfo {year} {2018})}\BibitemShut {NoStop}%
\bibitem [{\citenamefont {Flick}\ \emph {et~al.}(2019)\citenamefont {Flick},
  \citenamefont {Welakuh}, \citenamefont {Ruggenthaler}, \citenamefont
  {Appel},\ and\ \citenamefont {Rubio}}]{flick2019light}%
  \BibitemOpen
  \bibfield  {author} {\bibinfo {author} {\bibfnamefont {J.}~\bibnamefont
  {Flick}}, \bibinfo {author} {\bibfnamefont {D.~M.}\ \bibnamefont {Welakuh}},
  \bibinfo {author} {\bibfnamefont {M.}~\bibnamefont {Ruggenthaler}}, \bibinfo
  {author} {\bibfnamefont {H.}~\bibnamefont {Appel}},\ and\ \bibinfo {author}
  {\bibfnamefont {A.}~\bibnamefont {Rubio}},\ }\bibfield  {title} {\bibinfo
  {title} {Light--matter response in nonrelativistic quantum electrodynamics},\
  }\href@noop {} {\bibfield  {journal} {\bibinfo  {journal} {ACS photonics}\
  }\textbf {\bibinfo {volume} {6}},\ \bibinfo {pages} {2757} (\bibinfo {year}
  {2019})}\BibitemShut {NoStop}%
\bibitem [{\citenamefont {Mukamel}(1999)}]{mukamel1999principles}%
  \BibitemOpen
  \bibfield  {author} {\bibinfo {author} {\bibfnamefont {S.}~\bibnamefont
  {Mukamel}},\ }\href@noop {} {\emph {\bibinfo {title} {Principles of nonlinear
  optical spectroscopy}}},\ \bibinfo {number} {6}\ (\bibinfo  {publisher}
  {Oxford University Press on Demand},\ \bibinfo {year} {1999})\BibitemShut
  {NoStop}%
\bibitem [{\citenamefont {Iles-Smith}\ \emph {et~al.}(2017)\citenamefont
  {Iles-Smith}, \citenamefont {McCutcheon}, \citenamefont {Nazir},\ and\
  \citenamefont {M{\o}rk}}]{iles2017phonon}%
  \BibitemOpen
  \bibfield  {author} {\bibinfo {author} {\bibfnamefont {J.}~\bibnamefont
  {Iles-Smith}}, \bibinfo {author} {\bibfnamefont {D.~P.}\ \bibnamefont
  {McCutcheon}}, \bibinfo {author} {\bibfnamefont {A.}~\bibnamefont {Nazir}},\
  and\ \bibinfo {author} {\bibfnamefont {J.}~\bibnamefont {M{\o}rk}},\
  }\bibfield  {title} {\bibinfo {title} {Phonon scattering inhibits
  simultaneous near-unity efficiency and indistinguishability in semiconductor
  single-photon sources},\ }\href@noop {} {\bibfield  {journal} {\bibinfo
  {journal} {Nature Photonics}\ }\textbf {\bibinfo {volume} {11}},\ \bibinfo
  {pages} {521} (\bibinfo {year} {2017})}\BibitemShut {NoStop}%
\bibitem [{\citenamefont {Del~Pino}\ \emph {et~al.}(2018)\citenamefont
  {Del~Pino}, \citenamefont {Schr{\"o}der}, \citenamefont {Chin}, \citenamefont
  {Feist},\ and\ \citenamefont {Garcia-Vidal}}]{del2018tensor}%
  \BibitemOpen
  \bibfield  {author} {\bibinfo {author} {\bibfnamefont {J.}~\bibnamefont
  {Del~Pino}}, \bibinfo {author} {\bibfnamefont {F.~A.}\ \bibnamefont
  {Schr{\"o}der}}, \bibinfo {author} {\bibfnamefont {A.~W.}\ \bibnamefont
  {Chin}}, \bibinfo {author} {\bibfnamefont {J.}~\bibnamefont {Feist}},\ and\
  \bibinfo {author} {\bibfnamefont {F.~J.}\ \bibnamefont {Garcia-Vidal}},\
  }\bibfield  {title} {\bibinfo {title} {Tensor network simulation of
  non-markovian dynamics in organic polaritons},\ }\href@noop {} {\bibfield
  {journal} {\bibinfo  {journal} {Physical review letters}\ }\textbf {\bibinfo
  {volume} {121}},\ \bibinfo {pages} {227401} (\bibinfo {year}
  {2018})}\BibitemShut {NoStop}%
\bibitem [{\citenamefont {White}\ \emph {et~al.}(2021)\citenamefont {White},
  \citenamefont {Stewart}, \citenamefont {Solntsev}, \citenamefont {Li},
  \citenamefont {Toth}, \citenamefont {Kianinia},\ and\ \citenamefont
  {Aharonovich}}]{white2021phonon}%
  \BibitemOpen
  \bibfield  {author} {\bibinfo {author} {\bibfnamefont {S.}~\bibnamefont
  {White}}, \bibinfo {author} {\bibfnamefont {C.}~\bibnamefont {Stewart}},
  \bibinfo {author} {\bibfnamefont {A.~S.}\ \bibnamefont {Solntsev}}, \bibinfo
  {author} {\bibfnamefont {C.}~\bibnamefont {Li}}, \bibinfo {author}
  {\bibfnamefont {M.}~\bibnamefont {Toth}}, \bibinfo {author} {\bibfnamefont
  {M.}~\bibnamefont {Kianinia}},\ and\ \bibinfo {author} {\bibfnamefont
  {I.}~\bibnamefont {Aharonovich}},\ }\bibfield  {title} {\bibinfo {title}
  {Phonon dephasing and spectral diffusion of quantum emitters in hexagonal
  boron nitride},\ }\href@noop {} {\bibfield  {journal} {\bibinfo  {journal}
  {Optica}\ }\textbf {\bibinfo {volume} {8}},\ \bibinfo {pages} {1153}
  (\bibinfo {year} {2021})}\BibitemShut {NoStop}%
\end{thebibliography}%


%

\end{document}


\preprint{APS/123-QED}

\title{Supplementary Information: Signatures of Non-Markovianity in Cavity-QED with Color Centers in 2D Materials}
\author{Mark Kamper Svendsen}
\affiliation{CAMD, Department of Physics, Technical University of Denmark, 2800 Kgs. Lyngby,
Denmark}
\author{Sajid Ali}
\affiliation{CAMD, Department of Physics, Technical University of Denmark, 2800 Kgs. Lyngby,
Denmark}
\author{Nicolas Stenger}
\affiliation{Department of electrical and photonics engineering, Technical University of Denmark, 2800 Kgs. Lyngby, Denmark
}
\affiliation{
Center for Nanostructured Graphene, Technical University of Denmark, 2800 Kgs. Lyngby, Denmark
}
\affiliation{
NanoPhoton -- Center for Nanophotonics, Technical University of Denmark, 2800 Kgs. Lyngby, Denmark
}
\author{Kristian Sommer Thygesen}
\affiliation{CAMD, Department of Physics, Technical University of Denmark, 2800 Kgs. Lyngby,
Denmark}
\affiliation{
Center for Nanostructured Graphene, Technical University of Denmark, 2800 Kgs. Lyngby, Denmark
}
\author{Jake Iles-Smith}
 \affiliation{Department of Physics and Astronomy, The University of Manchester, Oxford Road, Manchester M13 9PL, United Kingdom}
  \affiliation{Department of Electrical and Electronic Engineering, The University of Manchester, Oxford Road, Manchester M13 9PL, United Kingdom}
\date{January 2022}

\maketitle

\section{Model and dynamical description}

In this section we will derive the dynamical description of an emitter interacting with electromagnetic and phonon environments used in the manuscript.
We start by considering the Hamiltonian of the emitter and environments, given by:
\begin{align}
    &H = H_\mathrm{S} + H_I^\mathrm{EM} + H_I^\mathrm{Ph}+ H_\mathrm{B}^\mathrm{EM} + H_\mathrm{B}^\mathrm{Ph}.
\end{align}
The time evolution of the emitter and the single mode the optical cavity is generated by the Jaynes-Cummings Hamiltonian $H_\mathrm{S}= \hbar\omega_e\sigma^\dagger\sigma + \hbar{}g(\sigma^\dagger a + \sigma a^\dagger) + \hbar\Omega_c a^\dagger a$, with the parameters defined in the manuscript.
In the dipole and rotating wave approximations, the interaction with the electromagnetic environment can be written as~\cite{carmichael1999statistical}:
\begin{equation}
    H_I^\mathrm{EM} = \sum_j \sum_\mathbf{l} (\hbar f_{j,\mathbf{l}} \hat{A}_j^\dagger c_{j,\mathbf{l}} + \hbar f^\ast_{j,\mathbf{l}} \hat{A}_j c^\dagger_{j,\mathbf{l}} ),
    \label{eq:EM_ham}
\end{equation}
where $\hat{A}_1=\sigma$ and $\hat{A}_2= a$, and $c_{j, \mathbf{l}}$ is the annihilation operator for the $\mathbf{l}^\mathrm{th}$~mode of the $j^\mathrm{th}$ electromagnetic environment and $f_{j,\mathbf{1}}$ the corresponding coupling strength.
The free evolution of the fields are then determined by $H_\mathrm{B}^\mathrm{EM} = \sum_{j,\mathbf{l}}\hbar\omega_{j,\mathbf{l}}c^\dagger_{j,\mathbf{l}}c_{j,\mathbf{l}}$.
The electron-phonon interaction is assumed to be linear~\cite{mahan2013many} and takes the form,
$$
   H_\mathrm{I}^\mathrm{Ph} = \sigma^\dagger\sigma\sum_{\mathbf{k}}\hbar g_\mathbf{k}(b_\mathbf{k}^\dagger + b_{-\mathbf{k}}),
$$
where the parameters are defined in the manuscript.
In the subsequent sections of this supplement, we will derive the dynamical description of system subject to these interactions, starting with the optical field, before introducing the Time Evolved Matrix Product Operator formalism (TEMPO)~\cite{strathearn2018efficient}.

\subsection{Coupling to the electromagnetic field}
To derive the impact of the electromagnetic fields, we first consider its influence in isolation from the electron-phonon interactions. 
By assuming the coupling between the cavity-emitter system and the external electromagnetic is weak, the resulting dynamics may be described by second order Born Markov master equation, which in the interaction picture takes the form~\cite{breuer2002theory, carmichael1999statistical}:
\begin{equation}
    \frac{\partial\rho(t)}{\partial t}=\mathcal{L}_\mathrm{EM}[\rho(t)] = \sum_{j}\sum_{\alpha\beta}\int_0^\infty d\tau \left(\left[S_{j,\alpha}(t),S_{j,\beta}(t-\tau)\rho(t)\right] C^{(j)}_{\alpha\beta}(\tau)+\mathrm{h.c.} 
    \right),
\end{equation}
where $S_{j,1} = A_j^\dagger$ and $S_{j,2} = A_j$ are the system coupling operators given in Eq.~\ref{eq:EM_ham} corresponding to the $j^\mathrm{th}$-optical environment. Similarly $C^{(j)}_{\alpha\beta} = \langle B_{j,\alpha}(\tau)B_{j,\beta}\rangle$ is the correlation function of the $j^\mathrm{th}$-environment, with $B_{j,1} = \sum_\mathbf{l} f_{j,\mathbf{l}}c_{j,\mathbf{l}}$ and $B_{j,2} = B_{j,1}^\dagger$. 
Taking the optical fields to be in their vacuum state, the only non-zero correlation function is
$
    C_{21}^{(j)}(\tau)=\sum_\mathbf{l}\vert f_{j,\mathbf{l}}\vert^2 e^{i\omega_{j,\mathbf{l}}\tau}
$. We can now make two further simplifications: the first is the flat spectrum approximation discussed in manuscript in which $f_{1,\mathbf{l}}\approx f_{1}= \sqrt{2\pi\Gamma}$ and $f_{2,\mathbf{l}}\approx f_{2}=\sqrt{2\pi\kappa}~\forall~\mathbf{l}$, which is valid in the weak coupling regime, and in situations in which the bath spectral density does not vary appreciably over the system energy scales. The second is to take the continuum limit, extending the lower limit of integration to $-\infty$~\cite{carmichael1999statistical}. In combination these approximations simplify the bath correlation functions to $\delta$-functions, such that $C^{(j)}_{21}(\tau)\approx \vert f_{j}\vert^2 \delta(\tau)/2\pi$.
We thus, end up with a master equation of the form:
\begin{equation}
    \mathcal{L}_\mathrm{EM}[\rho(t)] \approx \sum_j \frac{\vert f_j\vert^2}{4\pi}\left(2A_j\rho(t) A_j^\dagger - \{A_j^\dagger A_j, \rho(t)\}\right) = \Gamma L_{\sigma}[\rho(t)] + \kappa L_{a}[\rho(t)],
\end{equation}
where we have defined the lindblad dissipator $L_O[\rho] = O\rho O^\dagger - \{O^\dagger O, \rho\}/2$. Moving back into the Schrodinger picture, and including system Hamiltonian, we obtain $\mathcal{L}_\mathrm{0}\rho = -i[H_\mathrm{S},\rho] + \Gamma L_{\sigma}[\rho(t)] + \kappa L_{a}[\rho(t)]$.

\subsection{Deriving the influence functional and augmented density operator}
To derive the TEMPO algorithm, we follow Refs.~\cite{strathearn2018efficient}.
We start by considering the Liouville superoperator that generates the evolution of the system $\mathcal{L} = \mathcal{L}_0 + \mathcal{L}_E$, where $\mathcal{L}_0 = -i[H_\mathrm{S}, \cdot] + \mathcal{L}_\mathrm{EM}[\cdot]$ describes the dynamics of the emitter, while $\mathcal{L}_E = -i[H_\mathrm{I}^\mathrm{PH} + H_\mathrm{B}^\mathrm{PH},\boldsymbol{\cdot}]$ accounts for the dynamics of the phonon bath. 
The dynamics of this system are generated by the superoperator
$
    \mathcal{U}(t) = \exp(\mathcal{L}t)
$, such that the reduced state of the system is given by $\rho_\mathrm{S}(t) = \tr_E(\mathcal{U}(t)\rho_0\otimes\tau_B)$, where $\rho_0$ is the initial state of the system, and $\tau_B = \exp(\beta\sum_\mathbf{k}\nu_\mathbf{k} b_\mathbf{k}^\dagger b_\mathbf{k})/\mathcal{Z}$ is the Gibbs state of the phonon bath, with $\mathcal{Z}$ the corresponding partition function. 

We start by discretising the time interval, such that $t = k\delta t$ for integer $k$ and $\delta t \ll 1 $.
The propagator then becomes $\mathcal{U}(t) = \mathcal{U}_{\delta t}^N$, with $\mathcal{U}_{\delta t} = \exp(\mathcal{L}\delta t)$, allowing us to apply the symmetric Trotter splitting such that:
$
    \mathcal{U}_{\delta t} = \mathcal{V}_{\delta t}\mathcal{W}_{\delta t}\mathcal{V}_{\delta t} + \mathcal{O}(\delta t^3),
$
where we have introduced the superoperators $\mathcal{V}_{\delta t}[\rho] = \exp(\mathcal{L}_0 \delta t/2)[\rho]$, and $\mathcal{W}_{\delta t}[\rho] = U_{\delta t}\rho U_{\delta t}^\dagger$ with the unitary environment propagator defined as $U_{\delta t} = \exp(-i (H_I^\mathrm{PH} + H_B^\mathrm{PH})\delta t)$. 

For an initially uncorrelated system and environment, we can decompose the initial state of the system and environment into the system basis, such that $\chi(t=0) = \sum_{r_0s_0}\rho^{r_0s_0}\ket{r_0}\!\bra{s_0}\otimes \tau_B$, where $\rho^{s_0r_0} = \bra{r_0}\rho_0\ket{s_0}$. Applying the Trotterised propagator to the initial density operator yields:
\begin{equation}
    \mathcal{V}_{\delta t}\mathcal{W}_{\delta t} \mathcal{V}_{\delta t}[\ket{s_0}\!\bra{r_0}\otimes\tau] = \sum_{\alpha_1,\beta_1
    }\mathcal{V}_{\beta_1}^{\alpha_1}\mathcal{V}^{\beta_1}_{\alpha_0}\ket{r_1}\!\bra{s_1}\otimes\mathcal{W}^{\beta_1}[\tau_B],
\end{equation}
where we have introduced the compound indices $\alpha_i=(r_i, s_i)$ and $\beta_i = (t_i, u_i)$. By assuming the interaction term is diagonal in the system basis, the superoperators can be expanded such that $\mathcal{V}^{\alpha_i}_{\beta_i} = \bra{r_i}\mathcal{V}_{\delta t}[\ket{t_i}\bra{u_i}]\ket{s_i}$, and $\mathcal{W}^{\beta_i}[\tau] = \bra{t_i}\mathcal{W}_{\delta t}[\ket{t_i}\!\bra{u_i}\otimes\tau]\ket{u_i}$.
Applying this expression iteratively and tracing over the environmental degrees of freedom, we have after $k$ timesteps:
\begin{equation}
    \rho^{\alpha_k} = \sum_{\vec\alpha, \vec\beta} \rho^{\alpha_0}\prod_{j=1}^k\mathcal{V}_{\beta_{j}}^{\alpha_j}\mathcal{V}^{\beta_{j}}_{\alpha_{j-1}}\mathcal{F}_{\beta_{k}\cdots\beta_1},
    \label{eq:reduced}
\end{equation}
where $\rho^{\alpha_k} = \bra{r_{k}}\rho(k\delta t)\ket{s_k}$ and we have defined the influence functional as $\mathcal{F}_{\beta_k\cdots\beta_1}=\tr_E(\mathcal{W}^{\beta_k}\dots\mathcal{W}^{\beta_1}[\tau_B])$.
For an environment in a Gibbs state, the trace in the expression for the influence functional can be done analytically~\cite{makri1995tensorI,makri1995tensorII}, and written in product form as,
\begin{equation}
    \mathcal{F}_{\beta_{N}\cdots\beta_1} = \prod_{j=1}^N\prod_{ k=0}^{j-1}[b_{k}]_{\beta_j\beta_{j-k}},
\end{equation}
where we have defined the influence tensors:
\begin{equation}
    [b_{(i-j)}]_{\beta_i\beta_j} = e^{-(\lambda_{s_i}-\lambda_{r_i})(\eta_{i-j} \lambda_{s_j} - \eta_{i-j}^\ast\lambda_{r_j})}.
\end{equation}
Here $\lambda_{s}$ are the diagonal elements of the coupling operator, in this case $\sigma^\dagger\sigma$, and we have introduced the discretised memory kernel
\begin{equation}
    \eta_{i-j} = \left\{
    \begin{array}{cc}
    \int_{t_{i-1}}^{t_i}\int_{t_{j-1}}^{t_j} dt^\prime dt^{\prime\prime} C(t^\prime - t^{\prime\prime}), & ~\text{for}~i\neq j,\\
    \int_{t_{i-1}}^{t_i}\int_{t_{i-1}}^{t_i} dt^\prime dt^{\prime\prime} C(t^\prime - t^{\prime\prime}), & ~\text{for}~i=j.
    \end{array}
    \right.,
\end{equation}
with bath autocorrelation function,
\begin{align}
    C(t) = \int_0^\infty J_\mathrm{Ph}(\omega)\left[\mathrm{coth}\left(\frac{\beta \omega}{2}\right)\mathrm{cos}(\omega t) + i\mathrm{sin}(\omega t)\right].
\end{align}

It is clear from Eq.~\ref{eq:reduced}, that to calculate the state of the system at some time step $k$, requires the construction of the $(2k+1)$-index object:
\begin{equation}\begin{split}
    \mathcal{R}_{\beta_{k}\cdots\beta_1}^{\alpha_k\cdots\alpha_0} 
    =&\rho^{\alpha_0}\prod_{j=1}^{k}
    \mathcal{V}_{\beta_{j}}^{\alpha_j}
    \prod_{l=1}^{j}[{b}_{(j-l)}]_{\beta_j\beta_l}\mathcal{V}^{\beta_{j}}_{\alpha_{j-1}},
    \end{split}
    \label{eq:ADT}
\end{equation}
this object is referred to as the augmented density tensor (ADT), and, for arbitrarily small timesteps $\delta t$, is an exact representation of the system-environment interaction for the open quantum system of interest. 

The size of the ADT scales exponentially with number of timesteps taken. To circumvent this scaling, we follow Strathearn \emph{et al}~\cite{strathearn2018efficient} and represent the the ADT in matrix product operator form, utilising standard tensor compression methods~\cite{orus14intro} to reduce the rank of the elements in the tensor network.
We first start by considering the product of influence tensors in Eq.~\ref{eq:ADT}, $B^{\beta_k\cdots\beta_1}=\prod_{l=1}^{j}\left[b_{(j-l)}\right]_{\beta_j\beta_l}$. 
We can write this tensor in matrix product operator (MPO) form by defining the higher-rank tensor:
\begin{equation}
    \mathcal{B}_{~~\mu_{k-1}\cdots\mu_{1}}^{\beta_k \cdots\beta_1} = [b_0]_{\gamma_1}^{\beta_k}\left(\prod_{j=1}^{n-2}[b_{j}]^{\gamma_j, \beta_{k-j}}_{\gamma_{j+1},\mu_{k-j}}\right)[b_{k-1}]_{\mu_1}^{\gamma_{k-1},\beta_1},
\end{equation}
where we have promoted the influence tensors to rank 4 tensors using the Kronecker-$\delta$:
\begin{equation}
    [{b}_{(i-j)}]_{\beta_i\mu}^{\beta_j\gamma}= \delta_{\beta_i}^\gamma\delta_\mu^{\beta_j}[{b}_{(i-j)}]_{\beta_i}^{\beta_j}.
\end{equation}
As discussed in~\cite{strathearn2018efficient} and~\cite{Jorgensen19causal}, each time step corresponds to an application of the MPO $\mathcal{B}$, which is represented as a tensor network. 
Each node in the network consists of the rank-4 influence tensors defined above. 
We then compress the network after each timestep using a singular value decomposition (SVD)~\cite{orus14intro} at each node, and make a low rank approximation, whereby singular values less than a threshold value $\varepsilon_\mathrm{C}$ are discarded. 
This reduces the internal bond dimension of the MPS, providing a controllable way to reduce the size of the tensors. 
Alongside the size of the timestep $d t$, the singular values cut-off will be a principle convergence parameter, which will be discussed in Section~\ref{sec:convergence}.

\section{Linear response theory with TEMPO}

In this section we extend TEMPO to calculate linear response functions, and subsequently the absorption spectrum. 
Following the example of Flick~\emph{et al.}~\cite{flick2019light}, we consider a system weakly perturbed by by a driving field, $H^\prime=H + H_D(t)$, where $H_D(t)=E(t)\cdot\mu$.
The above Hamiltonian can represent two distinct physical scenarios: probing the emitter degrees of freedom using a classical driving field $E(t)$, or probing the cavity fields with $E(t)$ representing an external current.
In the two different cases, the system transition operator take on two values, $\mu=\sigma^\dagger + \sigma,\mathrm{\, or\, } a^\dagger + a$ respectively. 
The first is the dipole operator of the defect, describing the scattering of light directly off the two level transition, and forms the basis of standard linear response theory~\cite{mukamel1999principles}. 
The second is the quadrature operator of the cavity field, which induces a static polarisation of the cavity mode and excites real photons into the field~\cite{flick2019light}. Conceptually we can separate these two situations into different modes of driving, where the external field perturbs the cavity mode or the two level transition directly.

Treating the driving as weak, 
we follow the density matrix perturbation theory formalism outlined by Mukamel~\cite{mukamel1999principles}, and treat $H_D(t)$ perturbatively. Moving into the interaction picture with respect to the global Hamiltonian $H$, the density operator may be written as,
\begin{equation}
    \chi_\mathrm{I}(t) = U^\dagger(t,t_0)\chi(t)U(t,t_0),
\end{equation}
where $\chi$ represents a state in the full Hilbert space of the open quantum system, and $U(t,t_0) = \exp\left[-i H (t-t_0)\right]$.
The equation of motion for the state is then governed by the interaction picture von-Neumann equation
$
\dot{\chi}_\mathrm{I}(t) = -i\left[H_\mathrm{D}(t),\chi_\mathrm{I}(t)\right]
$.
We can formally write the solution to this equation in terms of the series expansion $\chi(t)= \chi^{(0)}(-\infty) + \sum_{n=1}^\infty\chi^{(n)}(t)$, where in the Schr\"odinger picture, the $n^\mathrm{th}$-order contribution takes the form
\begin{equation}
\chi^{(n)}(t) = (-i)^n\int\limits_{-\infty}^t d\tau_n\int\limits_{-\infty}^{\tau_n}d\tau_{n-1}\cdots 
\int\limits_{-\infty}^{\tau_2}d\tau_{1}U(t,t_0)
[H_D(\tau_n),[H_D(\tau_{n-1}),\cdots[H_\mathrm{D}(\tau_1),\chi(-\infty)]\cdots]U^\dagger(t,t_0).
\end{equation}
We choose the initial state of the open quantum system $\chi(-\infty)$, to be the equilibrium state such that it does not evolve in time.
Using this series expansion, we can consider the consider the effect the peturbative driving has on the polarisation of the field $P(t) = \tr(\mu\chi(t))=\sum_{n=0}^\infty P^{(n)}(t)$. By substituting in the expression for the driving Hamiltonian $H_D(t)$, and making a change of variables for the integrals, we obtain 
\begin{equation}\begin{split}
    P^{(n)} = \int_{0}^\infty dt_n\int_{0}^\infty dt_{n-1}\cdots 
\int_0^{\infty}dt_{1}
 E(t-t_n)E(t-t_n-t_{n-1})\cdots E(t-t_n\cdots-t_1)
 S^{(n)}(t_n,t_{n-1},\cdots, t_1),
\end{split}
\end{equation}
where we have introduced $n^\mathrm{th}$-order response function,
\begin{equation}
    S^{(n)}(t_n,\cdots,t_1) = (-i)^n\langle \mu(t_n + \cdots+t_1)[\mu(t_{n-1} + \cdots+t_1),\cdots[\mu(0),\chi(-\infty)]\cdots]\rangle.
\end{equation}
For this work, we will focus on the linear response of the system, that is, the first order response function:
\begin{equation}\label{eq:s1}
\begin{split}
    S^{(1)}(t) = -i\langle \mu(t)[\mu(0), \chi(-\infty)]\rangle=-i\left(\langle\mu(t)\mu(0)\chi(-\infty)\rangle -  \langle\mu(t)\mu(0)\chi(-\infty)\rangle^\ast\right).
\end{split}\end{equation}
For a defect coupled to phonons and an optical field, the equilibrium state is simply $\chi(-\infty) = \ket{g,0}\!\bra{g,0}\otimes\tau_B\otimes\ket{\{0\}}\!\bra{\{0\}}$, that is a product state of the system in its ground state, the phonons in Gibbs state $\tau_B$, and the electromagnetic field in the vacuum $\ket{\{0\}}=\bigotimes_\mathbf{l}\ket{0_\mathbf{l
}}$.
The trace in Eq. \ref{eq:s1} with the global equilibrium state $\chi(-\infty)$ means that even in the case of the bosonic quadrature operator $\mu = a + a^\dagger$  we never leave the single excitation Hilbert space in the case of linear response. 
We can therefore consider the linear response of the coupled cavity-TLE system within the single excitation subspace without loss of generality.

To obtain the linear spectrum we make a further simplification by working in the semi-impulsive limit, that is, we assume $E(t)\approx E_0 e^{i\omega_D t}\delta(t)$, where $E_0$ and $\omega_D$ are respectively the amplitude and frequency of the driving field.
In this case, the linear polarisation takes the simple form $P^{(1)}(t) = E_0 S^{(1)}(t)$, and we can link this to the field scattered by the system simply as $E_\mathrm{Out}(t) \propto i P^{(1)}(t)$.
The general expression for the (heterodyne) linear absorption spectrum can be found as~\cite{mukamel1999principles}:
\begin{equation}
    A(\omega)= 2\mathrm{Re}\left[\int_0^\infty dt e^{i\omega t}\tr\left(\mu(t)\mu(0)\chi(-\infty)\right)\right].
\end{equation}
Therefore, to obtain the linear spectrum, we simply need to calculate $\tr\left(\mu(t)\mu(0)\chi(-\infty)\right)$. This can be done using TEMPO by propagating a system in the unphysical initial state $\varrho(0) = \mu(0)\ket{g,0}\!\bra{g,0}$, and tracing with relevant driving operator $\mu$ at time $t$. Formally, one would have to propagate the system to $t = \infty$. However, in practice the important thing to ensure is that the system propagated to equilibrium.

\section{\emph{ab initio} spectral densities and \ctcn calculations}

In addition to the \cbvn defect, we also investigated the effect of the cavity on the \ctcn defect complex, which is considered another strong candidate as the source of single photon emission in hBN. In this section, we outline how we extract the spectral densities from the output of the atomistic simulations, before presenting the results for \ctcn\!. Note that we shall not give a detailed description of the atomistic simulation, focusing only on the aspects important to the manuscript.

\subsection{Spectral function of the electron phonon coupling}
As discussed in the manuscript, the electron phonon coupling of the defect complexes studied are treated from first-principles using density function theory (DFT). Specifically, the normal modes of the lattice are calculated independently for the defect complex in its ground and excited state configuration. Partial Huang Rhys parameters, $S_\mathbf{k}$, are then calculated using the Duschinsky method~\cite{borrelli2013franck}, which accounts for the possible geometric differences of phonon modes in the ground and excited state configurations. 

We can define the spectral density of the electron phonon coupling in terms of these partial Huang Rhys parameters, $\mathcal{J}_\mathrm{Ph}(\nu)= \sum_\mathbf{k}S_\mathbf{k}\delta(\nu-\nu_k)$. To account for the finite lifetime of these phonon modes, and to obtain a smooth function for the spectral density, we approximate the $\delta$-function with a Gaussian, such that $\mathcal{J}_\mathrm{Ph}(\nu) = \sum_\mathbf{k}S_\mathbf{k} f(\nu-\nu_\mathbf{k})$, where we have introduced the broadening function $f(\nu-\nu_\mathbf{k})=\frac{1}{\sqrt{2\pi}\sigma}\exp[-(\nu-\nu_k)^2/2\varsigma^2]$, with broadening parameter $\varsigma$. $\varsigma$ was chosen to 2.5 meV for \cbvn and 10 meV for \ctcn, similar to the values used in Refs. \cite{sajid2020vncb} and \cite{li2021c_2c_n} respectively.

\begin{figure}[htb]
    \centering
    \includegraphics[width=0.7\textwidth]{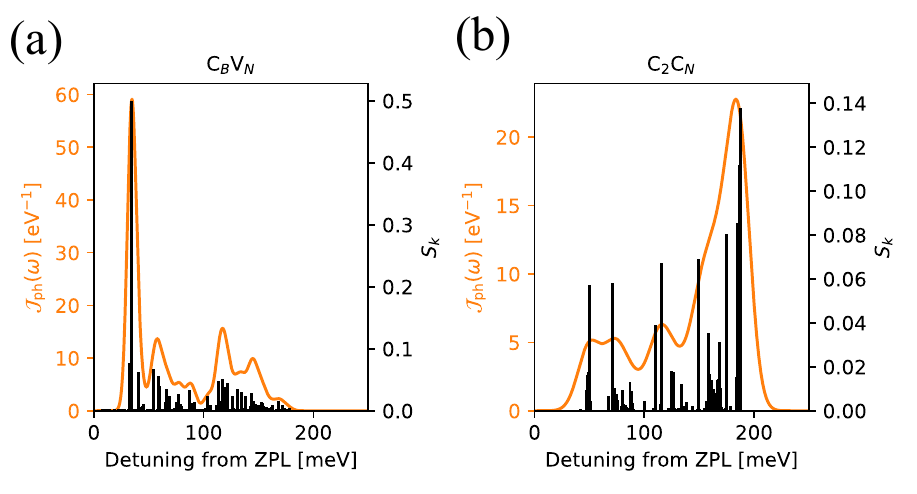}
    \caption{The partial Huang-Rhys parameters and broadened spectral functions for both \cbvn~and \ctcn.}
    \label{si_fig:partial_HRF}
\end{figure}

Fig. \ref{si_fig:partial_HRF} shows the partial Huang-Rhys factors in black, and the broadened spectral function of the electron-phonon coupling in orange, for the \cbvn (a) and \ctcn (b) defect complexes respectively. In contrast to the \cbvn spectral density, the phonon environment of C$_2$C$_N$ has very little structure at low energies, with a dominant contribution at $\hbar\omega\approx187$~meV. This mode is an out-of-phase phonon mode and discussed in Ref.~\cite{li2021c_2c_n}.

\subsection{Results for the C$_2$C$_N$ defect}

\begin{figure}[htb]
    \centering
    \includegraphics[width=\textwidth]{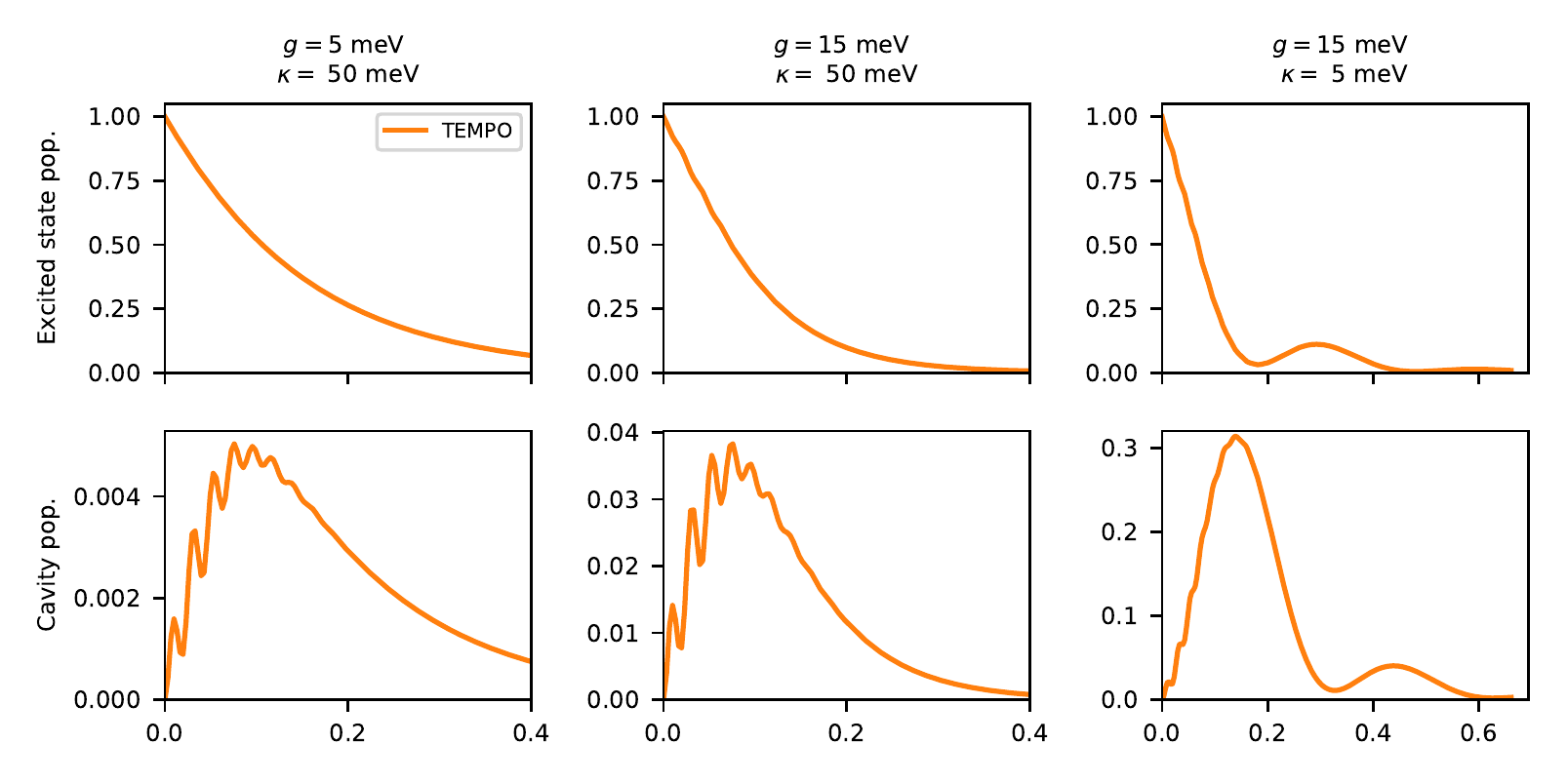}
    \caption{Time dependent emitter population (top) and cavity occupation (bottom) for different coupling strengths and cavity widths for the \ctcn~defect. Parameters used are $T=4$~K, $\Gamma=4$~meV, and $\hbar\Omega_c =\hbar \omega_e - \hbar\lambda$ where $\lambda=\int_0^\infty d\omega \omega^{-2}\mathcal{J}_\mathrm{Ph}(\omega)$ is the reorganisation energy. The step size used to obtain convergence was $dt=3.2$ fs, with SVD cut-off $\epsilon_\mathrm{C}=10^{-6}$, the same as for \cbvn~in the main manuscript.}
    \label{si_fig:c2cn_dynamics}
\end{figure}

Fig~\ref{si_fig:c2cn_dynamics} shows the dynamics of the emitter population and cavity occupation of the \ctcn~defect complex as a function of time for the same cavity parameters as considered for \cbvn~in the main manuscript. We clearly observe some structure in the population dynamics, but it is clear that the lack of a dominant phonon peak at small detuning from the ZPL means that the population dynamics of the \ctcn~defect are significantly less affected by the non-Markovian effects than was the case for \cbvn.

\begin{figure}
    \centering
    \includegraphics[width=0.7\textwidth]{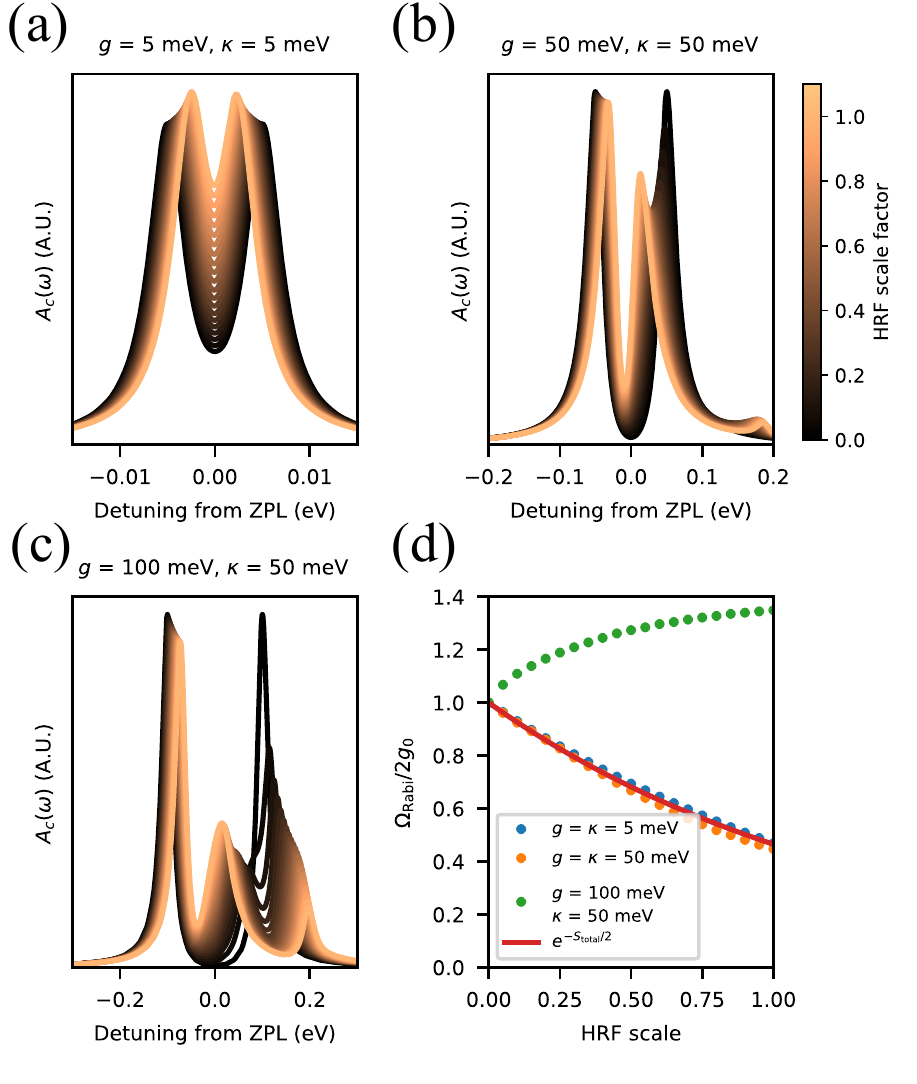}
    \caption{Effect of the electron-phonon coupling on the interaction with the cavity mode for the C$_2$C$_N$ defect. (a-c) the linear absorption spectra when probing the cavity mode, for various values of the scaling parameter $\alpha$. Here structure in the absorption spectra becomes increasingly important for large coupling, and broad cavity widths, but the effect is less pronounced than for \cbvn. (d) shows the change in peak position as a function of $\alpha$ (points) for the cavity parameters studied in (a-c)}
    \label{si_fig:c2cn_spectrum}
\end{figure}

Following the procedure in the manuscript, Figs.~\ref{si_fig:c2cn_spectrum} (a-c) show the absorption spectra when an artificial scaling of the Huang Rhys factor (HRF), $\alpha_\mathrm{HRF}$, is introduced to the phonon spectral density for \ctcn. 
Again, we see that the impact of phonons for $g=50$~meV is reduced in comparison to those found in the example of \cbvn.  
This is emphasised in Fig.~\ref{si_fig:c2cn_spectrum}~(d), where we plot the renormalised Rabi spltting as a function of the HRF scaling parameter. 
Here we see that for $\hbar{}g=5$~meV and $\hbar{}g=50$~meV the scaling matches the trend predicted by the Frank Condon factor $\mathscr{F}$, defined in the manuscript.
We do, however, see clearly the emergence of the hybridised polaron-polariton state in Fig.~\ref{si_fig:c2cn_spectrum}~(c), which is reflected in the departure from Frank Condon physics for $\hbar g=100$~meV and $\hbar\kappa=50$~meV, where the polariton splitting approaches the high-$Q$ mode of \ctcn~at $~\sim100$~meV.

\section{Computational details of the TEMPO calculations}
\label{sec:convergence}
The two parameters that need to be converged in TEMPO calculations are the time step and the singular value decomposition cut-off used in the truncation of the bond dimension of the MPS. For all configurations considered in this work, we find that a SVD cut-off of $\lambda_c = 10^{-6}$ is sufficient to ensure converged calculations. For each set of parameters, we separately converge the time step and we end up using a time step of $\delta t=5$~eV$^{-1}$ for the simulations with $g = 5,~15$ meV, and $\delta t=3$ eV$^{-1}$ and 2 eV$^{-1}$ for $g=50$ meV and 100 meV respectively.

For the spectrum calculations, it is important to ensure that the system is propagated long enough to capture the full dynamics, which we ensure in our calculations. However, because the frequency space resolution after the FFT is inversely proportional to the number of elements in the time array, it makes sense to pad with zeros to improve the resolution of the spectrum. We emphasize that since we ensure that all dynamics is properly accounted for by propagating the system to equilibrium, this padding has no effect on the physics of the system. We pad all time arrays such that the full array includes $2^{15}$ time steps. Note that it is crucial that the total number of array elements is a power of 2 to ensure that the FFT needed to calculate the spectrum runs efficiently.

\clearpage
\bibliography{references}